\documentclass[aps,prl,twocolumn,superscriptaddress,showpacs]{revtex4}
\usepackage[colorlinks=true,urlcolor=blue,citecolor=blue,linkcolor=blue]{hyperref}
\usepackage{graphicx}
\usepackage{mathrsfs}
\usepackage{bm}
\usepackage{verbatim}
\usepackage{latexsym}
\usepackage{stmaryrd}
\usepackage{amsmath}
\usepackage{amssymb}
\usepackage{epstopdf}
\usepackage{slashed}
\usepackage{bm}
\usepackage{hyperref}
\usepackage{chngcntr}
\usepackage{makecell}
\usepackage{footmisc}
\usepackage{diagbox}
\newcommand{\B}[1]{{\textcolor{blue}{#1}}}

\begin{document}
\title{ Bulk-Bulk Correspondence in Disordered Non-Hermitian Systems}

\author{Zhi-Qiang Zhang}
\affiliation{School of Physical Science and Technology, Soochow University, Suzhou 215006, China}
\affiliation{Institute for Advanced Study, Soochow University, Suzhou 215006, China}
\author{Hongfang Liu}
\affiliation{School of Physical Science and Technology, Soochow University, Suzhou 215006, China}
\affiliation{Institute for Advanced Study, Soochow University, Suzhou 215006, China}
\author{Haiwen Liu}
\affiliation{Center for Advanced Quantum Studies, Department of Physics, Beijing Normal University, Beijing 100875, People's Republic of China}
\author{Hua Jiang} \email{jianghuaphy@suda.edu.cn}
\affiliation{School of Physical Science and Technology, Soochow University, Suzhou 215006, China}
\affiliation{Institute for Advanced Study, Soochow University, Suzhou 215006, China}
\author{X. C. Xie}
\affiliation{International Center for Quantum Materials, School of Physics, Peking University, Beijing 100871, China}
\affiliation{CAS Center for Excellence in Topological Quantum Computation, University of Chinese Academy of Sciences, Beijing 100190, China}

\date{\today}
\begin{abstract}
The consistency between eigenvalues calculated under open and periodic boundary conditions, named as {\it bulk-bulk correspondence} ($\mathcal{BBC}$), can be destroyed in systems with non-Hermitian skin effect (NHSE). In spite of the great success of the generalized Brillouin zone (GBZ) theory in clean non-Hermitian systems,
the applicability of GBZ theory is questionable when the translational symmetry is broken. Thus, it is of great value to rebuild the $\mathcal{BBC}$ for disorder samples, which extends the application of GBZ theory in non-Hermitian systems.
Here, we propose a scheme reconstructing $\mathcal{BBC}$, which can be regarded as the solution of an optimization problem.
By solving the optimization problem analytically, we reconstruct the $\mathcal{BBC}$ and obtain the modified GBZ theory in several prototypical disordered non-Hermitian models. The modified GBZ theory gives a precise description of the fantastic NHSE, which predicts
two coupled disordered Hatano-Nelson chains with right-propagating NHSEs resulting in left-propagating NHSEs.
\end{abstract}

\date{\today}
\maketitle

%%%%%%%%%%%%%%%%%%%%%%%%%%%%%%%

\textit{ Introduction.---}
In recent years, non-Hermitian systems have been attracting great interests \cite{NH1,NH2,NH3,NH4,NH5,NH7,NH8,NH9,NH10,CS1,CS2,CS3,CS4,CS5,CS6,CS7,SL1,SL2,SL3,SL4,SL5,SL6,SL7,ref1,ref2,ref3,ref4,ref5,ref6,ref7,ref8,ref9,ref10,ref11,ref12,ref13,ref14,ref15,ref16,ref17,ref18,ref19,ref20,ref21,ref22,ref26,WER,XiongY,ref23,ref24,NHHOTI_ex,ref25}, among which the non-Hermitian skin effect (NHSE) is one of the most focused ones \cite{XiongY,ref23,ref24,ref25,NHSE1,NHSE2,NH6,NHSE3,NHSE4,NHSE5,NHSE6,NHSE7,NHSE8,NHSE9,NHSE10,NHSE11,NHSE12,NHSE13,NHSE14,NHSE15,NHSE16,NHSE17,NHSE18,NHSE19,NHSE20,NHSE21,NHSE22,NHSE23,NHSE24,NHHOTI_ex,NHSE25,NHSE26,Fsong}. The presence of NHSE indicates that the eigenvectors exhibit different distributions for open ($\mathcal{OBC}$) and periodic boundary conditions ($\mathcal{PBC}$) \cite{NHSE1}, where the eigenvectors are localized at specific areas of the sample under $\mathcal{OBC}$.
Such unique features also lead to the failure \cite{BBC1,BBC2,BBC3,BBC4,BBC5} of the bulk-bulk  correspondences ($\mathcal{BBC}$) due to the inconsistency of the eigenvalues between $\mathcal{OBC}$ and $\mathcal{PBC}$ in non-Hermitian systems \cite{NH6,NHSE3,NHSE4}.
The savior is the establishment of the generalized Brillouin zone (GBZ) theory \cite{NHSE1,NHSE2,NH6}, by which the problems of $\mathcal{BBC}$ are dealt with appropriately.
The correct open-boundary bulk spectra are available under $\mathcal{PBC}$ \cite{NH6,NHSE3,NHSE4}.
The validity of $\mathcal{BBC}$ also provides a reliable way for the characterization of NHSE, thus, a precise description of NHSE is now available \cite{NHSE3}.
Enlightened by the GBZ theory, a blossom of studies are reported with the combinations of topology and non-Hermiticity \cite{ref1,ref2,ref3,ref4,ref5,ref6,ref7,ref8,ref9,ref10,ref11,ref12,ref13,ref14,ref15,ref16,ref17,ref18,ref19,ref20,ref21,ref22,ref26,WER,XiongY,ref23,ref24,NHHOTI_ex,ref25}.

Very recently, the study on disorder effect in non-Hermitian systems has also drawn extensive attentions \cite{dis0,dis1,dis2,dis3,dis4,dis5,dis6,dis7,dis8,dis9,dis10,disTaler,dis11,disorder12}.
Importantly, the NHSE could still exist in disordered samples \cite{CS3,CS4,CS5,CS6,SL3,SL6} and leads to the breakdown of $\mathcal{BBC}$ for translational-symmetry-broken samples.
Although has achieved great successes in clean systems,
the applicability of GBZ theory in disordered systems is still not fully understood because this theory is heavily based on the translational symmetry \cite{NHSE1,NHSE2,NH6}.
Therefore, the widely adopted GBZ theory may not be directly applicable \cite{Fsong} when disorder is presented \cite{dis0,dis1,dis2,dis3,dis4,dis5,dis6,dis7,dis8,dis9,dis10,disTaler,dis11,disorder12,HNM,Anderson,Anderson1,Anderson2}. The $\mathcal{BBC}$ mechanism as well as the quantitative description of the NHSE in disordered samples remain to be investigated.

 \begin{figure}[t]
\includegraphics[width=8cm]{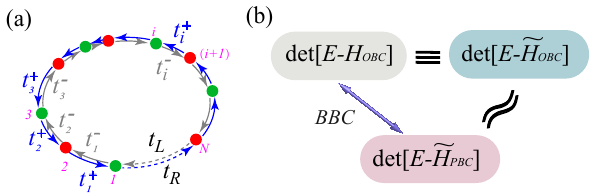}
\caption{ (a). Schematic diagram of the Hatano-Nelson model. The blue ($t_i^+$) and gray ($t_{i}^-$) arrows are the asymmetric hopping. The dashed lines correspond to the $\mathcal{PBC}$ with hopping strength $t_{L/R}$. (b) Key relationships between the determinants under $\mathcal{PBC}$ and $\mathcal{OBC}$ for the initial Hamiltonian $\mathcal{H}$ and the transformed Hamiltonian $\mathcal{\widetilde{H}}$ to obtain the $\mathcal{BBC}$. }
\label{f1}
\end{figure}

In this Letter, we propose that the universal scheme reconstructing $\mathcal{BBC}$ in disordered non-Hermitian systems can be reduced to an optimization problem in linear algebra.
By analytically solving this problem, we demonstrate two key issues about the $\mathcal{BBC}$: (i) A Hamiltonian preserves its determinant after a similarity transformation under $\mathcal{OBC}$; (ii) The transformed Hamiltonian possesses the $\mathcal{BBC}$.
As an illustration, we study the disorder effect on several
prototypical non-Hermitian models. We show that the GBZ theory in the clean limit cannot be directly applied to the disordered non-Hermitian samples.
However, the correct $\mathcal{BBC}$ can be recovered in the manner of the similarity transformation stated above.
In this way, we obtain a modified GBZ theory which extends the application of the GBZ theory into disordered systems.
Guided by this theory, we predict the emergence of disorder-enhanced and disorder-irrelevant NHSEs as well as coupling-reversed NHSE's directions.

\textit{ Model.---}
We start from the disordered Hatano-Nelson model \cite{HNM} shown in Fig. \ref{f1}(a) with Hamiltonian:
\begin{align}
\begin{split}
\mathcal{H} =\sum_{i=1}^{N} -t^+_ic_{i}^\dagger c_{i+1}-t^-_{i}c_{i+1}^\dagger c_{i}.
\label{EQ1}
\end{split}
\end{align}
We set $t^+_i=(t+\gamma+w^+_i)$ and $t^-_{i}=(t-\gamma+w^-_{i})$. $w^{\pm}_{i}\in[-\frac{W}{2},\frac{W}{2}]$ denotes the Anderson disorder \cite{Anderson,shiyan1} with $W$ the disorder strength. $t^\pm_i$ is the asymmetric hopping between the $i$-$th$ and the $(i+1)$-$th$ sites. Supposing that the $1st$ site overlaps with the $(N+1)$-$th$ site,
then one has the $\mathcal{PBC}$ ($\mathcal{OBC}$) if $t_R=t^+_{N}\equiv (t+\gamma+w^+_N)$ and $t_L=t^-_{N}\equiv (t-\gamma+w^-_{N})$ are nonzero (zero).

When $\gamma\neq0$ and $W=0$, the NHSE is presented in such a model. The GBZ theory should be adopted \cite{NHSE4} to achieve the $\mathcal{BBC}$.
For samples with a finite size, the GBZ theory introduces a transformation \cite{NHSE1,NHSE2,NH6}
\begin{align}
\begin{split}
\mathcal{H}(t^+_i,t^-_{i})\rightarrow\mathcal{\widetilde{H}}(\widetilde t^+_i,\widetilde t^-_{i}),
\label{EQ2}
\end{split}
\end{align}
where $\widetilde{t}^+_{i}=\beta_it^+_i$, and $\widetilde{t}^-_{i}=\beta_i^{-1}t^-_{i}$. $\beta_i\equiv \beta_{\mathcal{GBZ}}=\sqrt{\frac{t-\gamma}{t+\gamma}}$ modulates the Bloch wavefunction.
 However, the derivation of $\beta_{\mathcal{GBZ}}$ relies on the translational symmetry \cite{NHSE1,NH6}.
 When the disorder effect is included, the $\beta_i$ cannot be obtained by the previous version of the GBZ theory.

\textit{ Proposed scheme reconstructing $\mathcal{BBC}$.---}
We find that $\mathcal{BBC}$ in disordered non-Hermitian systems can be reconstructed as follows, where the translational symmetry is dispensable.
Concretely, the construction of $\mathcal{BBC}$ for samples with NHSE relies on the following two issues: (i) {\it A similarity transformation $\mathcal{H} \to \mathcal{\widetilde{H}}$ should preserve the determinant under $\mathcal{OBC}$} as $\det[E-\mathcal{H_{OBC}}]\equiv\det[E-\mathcal{\widetilde{H}_{OBC}}]$; (ii) {\it Such a similarity transformation should eliminate the NHSE so the transformed Hamiltonian $\mathcal{\widetilde{H}}$ possesses the $\mathcal{BBC}$ as $\det[E-\mathcal{\widetilde{H}_{OBC}}]
\approx\det[E-\mathcal{\widetilde{H}_{PBC}}]$}.
These relations can be summarized in a schematic diagram shown in Fig. \ref{f1}(b), which can be formulated as
\begin{align}
\begin{split}
\det[E-\mathcal{H_{OBC}}]\equiv\det[E-\mathcal{\widetilde{H}_{OBC}}]
\approx\det[E-\mathcal{\widetilde{H}_{PBC}}].
\label{EQ4}
\end{split}
\end{align}
\noindent Equation (\ref{EQ4}) leads to the $\mathcal{BBC}$ as the reason shown below.
Firstly, $E_{\mathcal{OBC}}$, the eigenvalue of $\mathcal{H_{OBC}}$ satisfies $\det[E_{\mathcal{OBC}}-\mathcal{H_{OBC}}]=0$.
If Eq. (\ref{EQ4}) holds, then the eigenvalues of $\mathcal{\widetilde{H}_{PBC}}$ should be approximately the same as those in $\mathcal{H_{OBC}}$, i.e.,
$\det[E_{\mathcal{OBC}}-\mathcal{\widetilde{H}_{PBC}}]\approx0$.
Physically, Eq. (\ref{EQ4}) defines the criterion to achieve appropriate bulk states for non-Hermitian systems.
In this way, the $\mathcal{BBC}$ and a detailed description of NHSE for $\mathcal{H}$ are available. Notably, the validity of this criterion is model-independent since it is insensitive to the form of $\mathcal{H}$.
Figure \ref{f1}(b) sketches the key finding of this Letter, where an appropriate transformation is desired to match Eq. (\ref{EQ4}).

\textit{ BBC as an optimization problem.---}
Generally, the accurate eigenvalues under the $\mathcal{OBC}$ cannot be obtained under the $\mathcal{PBC}$.
Nevertheless, the $\mathcal{BBC}$ can still be captured by solving an optimization problem related to Eq. (\ref{EQ4}) as shown below.
Taking the disordered Hatano-Nelson model with sample size $N=n$ as an example, the determinant under $\mathcal{PBC}$ reads \cite{SM,LNAB}:
\begin{align}
\begin{split}
 \det[E-\mathcal{H}_{\mathcal{PBC};~n\times n}]=&\det[E-\mathcal{H}_{\mathcal{OBC};~n\times n}]+f_{\mathcal{PBC}},
\label{EQS4}
\end{split}
\end{align}
with
\begin{align}
\begin{split}
f_{\mathcal{PBC}}&=(-1)^{\tau}t_Lt_R\det[E-\mathcal{H}_{\mathcal{OBC};~n-2\times n-2}]\\
&+(-1)^{n+1}[\prod\limits_{i=1}^{n-1} t_Rt^+_i+\prod\limits_{i=1}^{n-1}t_Lt^-_{i}].
\end{split}
\end{align}
Here, $\tau\equiv\tau(n;2,\cdots,n-1;1)$ is the permutation of $n$-$th$ order \cite{SM,LNAB} [see Supplementary Material (SM). VII].
By considering $E_{\mathcal{OBC}}$ as one of the open boundary eigenvalues of $\mathcal{H}$ with dimension $(n-2)$, one has
$\det[E_{\mathcal{OBC}}-\mathcal{H}_{\mathcal{OBC};~n-2\times n-2}]=0$.
\noindent When $n\rightarrow \infty$, it is reasonable \cite{SM} to show
$\det[E_{\mathcal{OBC}}-\mathcal{H}_{\mathcal{OBC};~n\times n}]\approx0$, then
Eq. (\ref{EQS4}) can be rewritten as
\begin{align}
\begin{split}
|\det[E_{\mathcal{OBC}}-\mathcal{H}_{\mathcal{PBC}}]|\approx|[\prod_{i=1}^{n-1} t_Rt^+_i+\prod_{i=1}^{n-1}t_Lt^-_{i}]|.
\label{EQx}
\end{split}
\end{align}
\noindent Here $|\cdots|$ stands for modulus. The right hand side of Eq. (\ref{EQx}) is nonzero in general, hence the accurate value of $E_{\mathcal{OBC}}$ cannot be obtained under $\mathcal{PBC}$. Moreover, in the presence of NHSE, the asymmetric hopping could significantly increase \cite{SM} the modulus in Eq. (\ref{EQx}), which leads to the breakdown of $\mathcal{BBC}$.

We find $\det[E_{\mathcal{OBC}}-\mathcal{\widetilde{H}}_{\mathcal{PBC}}]$ gets closer to zero by inserting the transformation $\widetilde{t}^\pm_{i}=\beta^{\pm1}_it^\pm_i$ adopted in Eq. (\ref{EQ2}).
Importantly, such a transformation preserves the determinant under $\mathcal{OBC}$ as
$\det[E-\mathcal{\widetilde{H}_{OBC}}(\widetilde{t}^+_i,\widetilde{t}^-_{i})]=\det[E-\mathcal{H_{OBC}}(t^+_i,t^-_{i})]= f(t^+_it^-_{i},E)$ since $t^+_it^-_{i}=\widetilde t^+_i\widetilde t^-_{i}$ and
$f(t^+_it^-_{i},E)$ is a polynomial only depending on $t^+_it^-_{i}$ and $E$ \cite{SM}.
 Thus, the first issue for obtaining Eq. (\ref{EQ4}) is satisfied, and Eq. (\ref{EQS4}) can be rewritten as:
\begin{align}
\begin{split}
\mathcal{\widetilde{F}}\approx|[\prod_{i=1}^{n-1} \widetilde{t}_R\widetilde{t}^+_i+\prod_{i=1}^{n-1}\widetilde{t}_L\widetilde{t}^-_{i}]|,
\label{EQ8}
\end{split}
\end{align}
\noindent where $\mathcal{\widetilde{F}}\equiv|\det[E_{\mathcal{OBC}}
-\mathcal{\widetilde{H}}_{\mathcal{PBC}}]|$.
For a suitable $\beta_i$, $\mathcal{\widetilde{F}}$ reaches its minimum so the eigenvalues of $\mathcal{\widetilde{H}}$ under $\mathcal{PBC}$ ($\widetilde{E}_{\mathcal{PBC}}$) fit better with $E_{\mathcal{OBC}}$. This is equivalent to an optimization problem minimizing $|\widetilde{E}_{\mathcal{PBC}}-E_{\mathcal{OBC}}|$.

Specifically, minimizing $\mathcal{\widetilde{F}}$ requires $\beta_i\equiv\sqrt{t^-_{i}/t^+_i}$ [see SM. II
\cite{SM}]. When disorder is absent, it reduces to the GBZ theory $\beta_i\equiv\beta_{\mathcal{GBZ}}$ \cite{footnote5}.
For the Hermitian case without NHSE, Eq. (\ref{EQx}) is already minimized since $t^+_i=t^-_{i}$, which ensures that $\det[E-\mathcal{H_{OBC}}]\approx\det[E-\mathcal{H_{PBC}}]$ holds.
Similarly, $\beta_i\equiv\sqrt{t^-_{i}/t^+_i}$ leads to $\widetilde{t}^+_i=\widetilde{t}^-_{i}=\sqrt{t^+_it^-_{i}}$ so that $\mathcal{\widetilde{H}}$ has no asymmetric hopping. Thus, $\mathcal{\widetilde{H}}$ should have no NHSE and the $\mathcal{BBC}$ is rebuilt as $\det[E-\mathcal{\widetilde{H}_{OBC}}]\approx\det[E-\mathcal{\widetilde{H}_{PBC}}]$ \cite{footnote3}.

In short, for obtaining Eq. (\ref{EQ4}), one needs to manage to find a similarity transformation $\widetilde{t}^\pm_{i}=\beta^{\pm1}_it^\pm_i$ which minimizes $\mathcal{\widetilde{F}}$.
The $\beta_i$ is reduced to $\beta_{\mathcal{GBZ}}=\sqrt{(t-\gamma)/(t+\gamma)}$ in the clean limit by the GBZ theory.
 Due to the universality of Eq. (\ref{EQ4}) and its related results, one is able to study the $\mathcal{BBC}$ in disordered non-Hermitian systems.

\begin{figure}[t]
\includegraphics [width=8.2 cm]{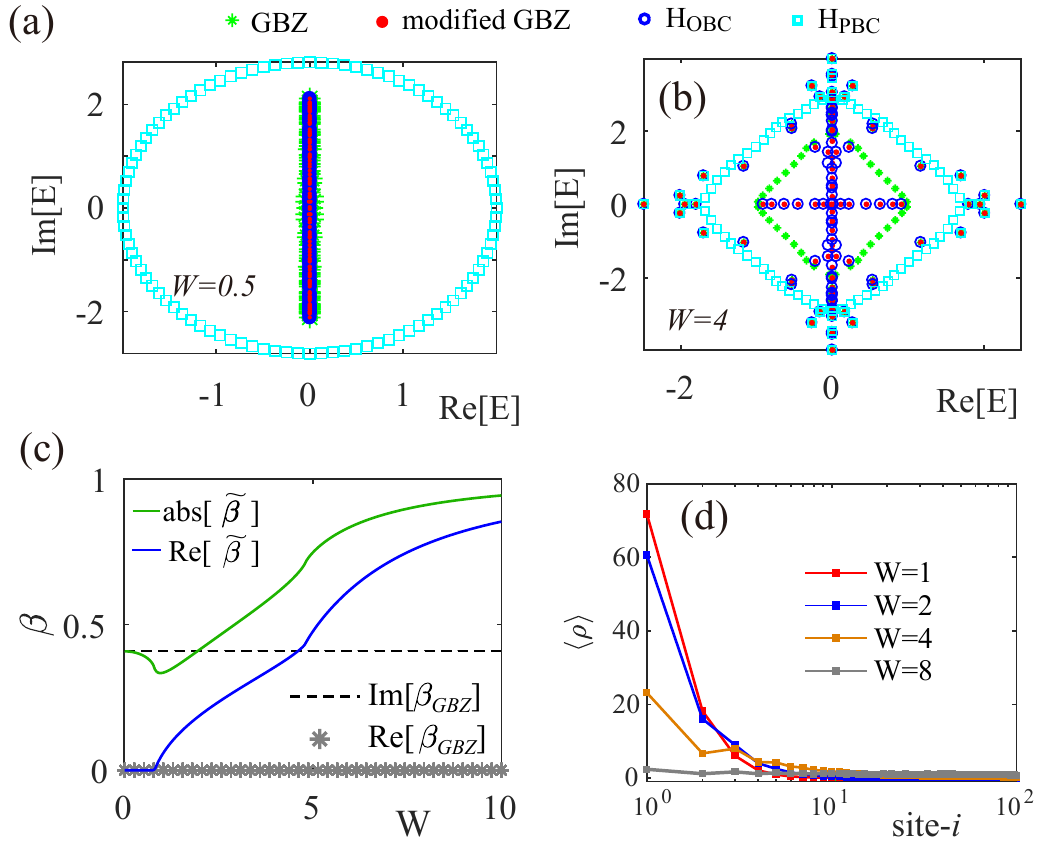}
\caption{ (a)-(b) The real part ($\mathrm{Re}[E]$) versus the imaginary part ($\mathrm{Im}[E]$) of the eigenvalues for different disorder strengths $W$ with $\gamma=1.4$ and sample size $N=100$.
 The GBZ are calculated under $\mathcal{PBC}$ with $\beta_i=\beta_{\mathcal{GBZ}}$.
  The modified GBZ are the same with GBZ, except  that $\beta_i$ is replaced by the transformation parameter $\widetilde{\beta}$ of the modified GBZ theory.
$\mathcal{H_{PBC}}$ and $\mathcal{H_{OBC}}$ are the original Hamiltonian $\mathcal{H}$ under $\mathcal{PBC}$ and $\mathcal{OBC}$, respectively.
   (c) A typical plot of $\beta$ versus $W$ with $\gamma=1.4$ for $\beta_{\mathcal{GBZ}}$ and $\widetilde{\beta}$.
 (d) The average density $\langle\rho\rangle=\langle\sum_{i\in \mathrm{all}} |\psi_i|^2\rangle$.}
\label{f2}
\end{figure}

\textit{Failure of $\beta_{\mathcal{GBZ}}$ in disordered samples.---}
To demonstrate the advantage of the proposed scheme reconstructing $\mathcal{BBC}$, we clarify that the NHSE is still presented for disordered samples, which is supported by the numerical results shown in Figs. \ref{f2}(a) and (b). Here the eigenvalues of $\mathcal{H_{PBC}}$ form a closed loop, which implies the presence of the NHSE \cite{NH10}. Furthermore, $\mathcal{H_{PBC}}$ and $\mathcal{H_{OBC}}$ show different eigenvalues, indicating that the $\mathcal{BBC}$ is destroyed.

Although no longer in the clean limit, for the sake of comparison, we can still try to utilize $\beta_{\mathcal{GBZ}}$ \cite{footnote5} to estimate the eigenvalues in disordered samples.
Based on Figs. \ref{f2}(a)-(b), one notices that $\beta_{\mathcal{GBZ}}$ gives the wrong (correct) open boundary spectrum under strong (weak) disorder.
Since $\beta_{\mathcal{GBZ}}$ preserves $\det[E-\mathcal{H_{OBC}}]=\det[E-\mathcal{\widetilde{H}_{OBC}}]$,
the failure of GBZ theory is attributed to the presence of
NHSE for $\mathcal{\widetilde{H}}$ with $\beta_i=\beta_{\mathcal{GBZ}}$
[see the closed loops in Fig. \ref{f2}(b)], which gives rise to $\det[E-\mathcal{\widetilde{H}_{PBC}}] \slashed{\approx} \det[E-\mathcal{\widetilde{H}_{OBC}}]$, i.e. violating Eq. (\ref{EQ4}).
 %Even though the GBZ theory gives correct results under weak disorder [see Fig. \ref{f2}(a)], strong disorder could significantly alter the NHSE.
 Therefore, for disordered samples, $\beta_{\mathcal{GBZ}}$ can not minimize $\mathcal{\widetilde{F}}$ [see SM. VII \cite{SM}] and the GBZ theory is no longer directly applicable.

\textit{ Modified GBZ theory.---}
Now we adopt the proposed scheme to rebuild the $\mathcal{BBC}$ for disordered samples. Although the transformation $\beta_i=\sqrt{t_i^-/t_i^+}$ minimizes $\widetilde{\mathcal{F}}$ \cite{SM}, $\beta_i$ differs from sample to sample when disorder is presented.
Now, we demonstrate that for a fixed disorder strength $W$, the $\beta_i$ in the modified GBZ theory can be replaced by a universal transformation parameter $\widetilde{\beta}$.
 Since $\widetilde{\beta}$ always preserves $\det[E-\mathcal{H_{OBC}}]=\det[E-\mathcal{\widetilde{H}_{OBC}}]$, one only needs to ensure that $\widetilde{\beta}$ preserves $\det[E-\mathcal{\widetilde{H}_{PBC}}]\approx\det[E-\mathcal{H_{OBC}}]$.
Based on Eq. (\ref{EQ8}), we find that the minimum \cite{SM} of
 $|\widetilde{\beta}^{N}t_R\prod_i t^+_i+\widetilde{\beta}^{-N}t_L\prod_{i} t^-_{i}|$
ensures that $\det[E-\mathcal{\widetilde{H}_{PBC}}]\approx\det[E-\mathcal{H_{OBC}}]$. Thus, one has
\begin{align}
\begin{split}
\widetilde{\beta}^{2N}\equiv\prod\limits_{i=1}^{N-1} \frac{t_Lt^-_{i}}{t_Rt^+_{i}}=\frac{\prod_{i=1}^{N} (t-\gamma+w^-_i)}{\prod_{i=1}^{N} (t+\gamma+w^+_i)}.
\label{modifidGBZx}
\end{split}
\end{align}
\noindent After some algebra including a self-average \cite{SM}, $\widetilde{\beta}$ for the modified GBZ theory with Anderson disorder reads:
\begin{equation}
\widetilde{\beta} = [ \frac{(t-\gamma+\frac{W}{2})^{t-\gamma+\frac{W}{2}}(t+\gamma-\frac{W}{2})^{t+\gamma-\frac{W}{2}}}
{(t-\gamma-\frac{W}{2})^{t-\gamma-\frac{W}{2}}(t+\gamma+\frac{W}{2})^{t+\gamma+\frac{W}{2}}} ] ^ {\frac{1}{2W}}.
\label{modifidGBZ}
\end{equation}
\noindent We set $\beta_i=\widetilde{\beta}$ here after \cite{footnote7}, which is a universal transformation parameter when $W$ is fixed.

The analytical results show that $\widetilde{\beta}$ could be complex as $\widetilde{\beta}=|\widetilde{\beta}|e^{j\theta}$ [see Fig. \ref{f2}(c)].
However, the modulus is more important for experimental observations.
In Fig. \ref{f2}(c), we plot the evolution of $|\widetilde{\beta}|$ versus $W$, and some specific features can be identified from both Eq. (\ref{modifidGBZ}) and the figure.
 In case of weak disorder, one has $\lim\limits_{W\rightarrow0}\widetilde{\beta}\sim\sqrt{\frac{t-\gamma}{t+\gamma}}$, which is the same with $\beta_{\mathcal{GBZ}}$.
 In case of strong disorder, $\lim\limits_{W\rightarrow\infty}|\widetilde{\beta}|\sim1$ implies the absence of NHSE with $\mathcal{H}\sim\mathcal{\widetilde{H}}$.
 Since $\beta$ should eliminate the NHSE of $\mathcal{H}$ to achieve the correct results, $|\widetilde{\beta}|$ can be considered as the strength of NHSE \cite{NHSE1,NH6}.
  Consequently, Fig. \ref{f2}(c) can be regarded as the phase diagram of NHSE, where $|\widetilde{\beta}|$ is analytically drawn from Eq. (\ref{modifidGBZ}). The NHSE is presented when $|\widetilde{\beta}|$ deviates from one that the more $|\widetilde{\beta}|$ deviates from one, the stronger the NHSE is.
 The direction of NHSE is determined by the sign of $|\widetilde{\beta}|-1$.
 More importantly, by adopting $\widetilde{\beta}$, as shown in Figs. \ref{f2}(a)-(b), the results for the modified GBZ theory under $\mathcal{PBC}$ (red dots) overlap perfectly with the eigenvalues for $\mathcal{H_{OBC}}$ (blue circles) \cite{footnote6}, verifying our theory.
Such verification is independent of the parameters \cite{SM}.

Furthermore, the advantage of the modified GBZ theory can be exhibited by comparing $|\beta_{\mathcal{GBZ}}|$ and $|\widetilde{\beta}|$, shown in Fig. \ref{f2}(c). $\beta_{\mathcal{GBZ}}$ overlaps with $\widetilde{\beta}$ in the modified GBZ under weak disorder, where the GBZ theory still holds [see Fig. \ref{f2}(a)].
For strong disorder, the modified GBZ captures the disorder-dependent NHSE, while $\beta_{\mathcal{GBZ}}$ fails. The disorder dependence of NHSE can be identified in Figs. \ref{f2}(a)-(b), where the eigenvalues of $\mathcal{H}_\mathcal{PBC}$ tend to lose their closed loop characteristics. It agrees with Fig. \ref{f2}(c), where $|\widetilde{\beta}|\rightarrow 1$ for strong disorder.
 Experimentally, the variation of NHSE can be detected by measuring the average density $\langle\rho\rangle=\sum_{i\in \mathrm{all}} \langle|\psi_i|^2\rangle$ where $\mathcal{H_{OBC}}\psi_i=E_i\psi_i$ \cite{NHSE14} and $\langle\cdots\rangle$ is the ensemble average. As plotted in Fig. \ref{f2}(d), the NHSE becomes weaker with the increase of the disorder strength. These numerical results are consistent with the analytical results of $|\widetilde{\beta}|$ shown in Fig. \ref{f2}(c).

\begin{figure}[t]
\includegraphics [width=8.4 cm]{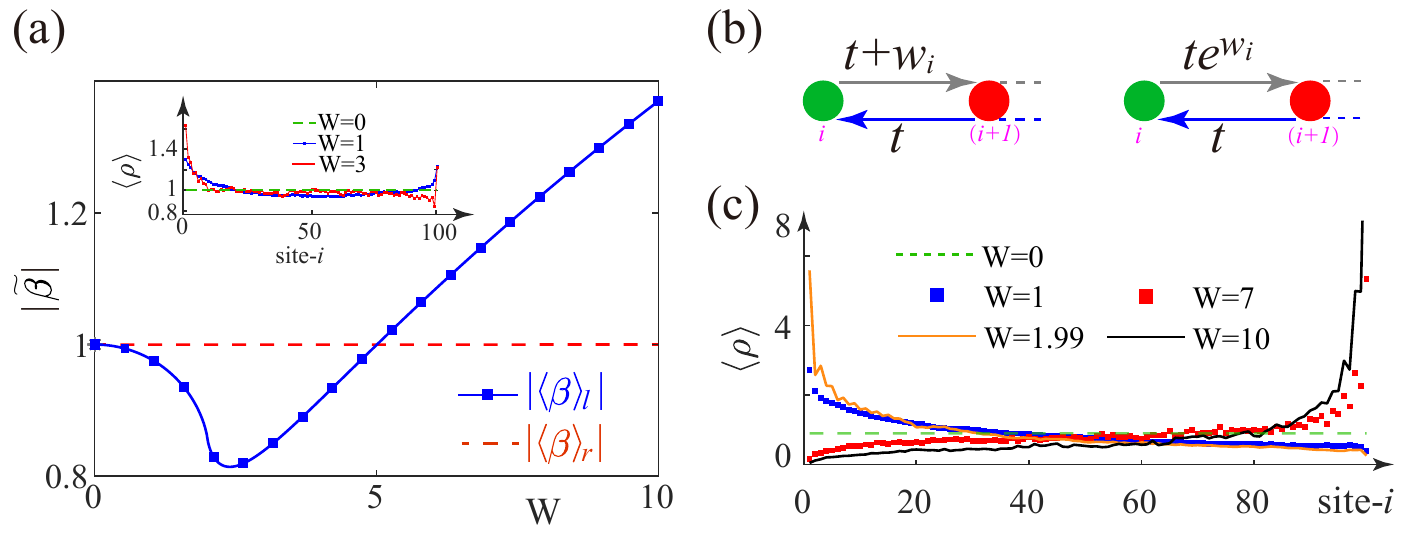}
\caption{(a) $|\widetilde{\beta}_{l/r}|$ versus disorder strength $W$. $|\widetilde{\beta}_{l}|$ ($|\widetilde{\beta}_{r}|$) corresponds to the left (right) panel in (b).
  Inset is the plot of eigenvectors $\langle\rho\rangle$ for the model in the right panel in (b) under different $W$.
 (b) Schematic diagrams of two typical disorder schemes.
 (c) $\langle\rho\rangle$ for model in the left panel in (b). }
\label{f3}
\end{figure}

\textit{Disorder-enhanced and disorder-irrelevant NHSEs.---}
The proposed theory deepens our understanding of $\mathcal{BBC}$ in disordered non-Hermitian systems.
As a demonstration, we predict the presence of disorder-enhanced and disorder-irrelevant NHSEs. Specifically, we start from a Hermitian Hamiltonian with $t^+_i=t^-_{i}=t$ and $t=1$ in Eq. (\ref{EQ1}). Such a model has no NHSE with $\beta_{\mathcal{GBZ}}\equiv1$. Two disorder schemes are considered as shown in Fig. \ref{f3}(b), where $t^-_{i}=t+w_i$ and $t^+_{i}=t$ in the left panel, while $t^-_{i}=te^{w_i}$ and $t^+_{i}=t$ in the right panel. The $w_i$ is uniformly distributed as $w_i\in[-\frac{W}{2},\frac{W}{2}]$.

At first glance, the above two models look the same, as the locally asymmetric hopping term is exerted.
However, these two disorder mechanisms exhibit distinct behaviors based on the criterion of $\mathcal{BBC}$.
The modified GBZ theory shows $\widetilde{\beta}_r=\beta_{\mathcal{GBZ}}\equiv1$, which is irrelevant with disorder strength, and the NHSE is absent shown in the inset of Fig. \ref{f3}(a) \cite{SM}. These phenomena manifest the disorder-irrelevant NHSE, where $\widetilde{\beta}\equiv \beta_{\mathcal{GBZ}}$ is irrelevant with $W$.
For model in the left panel of Fig. \ref{f3}(b),
a disorder-induced NHSE \cite{disTaler} is exhibited with $\widetilde{\beta}_l\equiv[\frac{(1+\frac{W}{2})^{1+\frac{W}{2}}e^{-W}}
{(1-\frac{W}{2})^{1-\frac{W}{2}}}]^{\frac{1}{2W}}$ [see Fig. \ref{f3}(a)].
Importantly, $|\widetilde{\beta}_l|$ precisely describes the variation of NHSE, where the disorder-enhanced NHSE is obtained. Such a feature can not be captured by self-consistent Born approximation \cite{footnote4}.
The plot of $\langle\rho\rangle$ is consistent with the variation of $|\widetilde{\beta}_l|$, where the disorder-enhanced NHSE is identified in Fig. \ref{f3}(c).
Effects of other types of disorder can also be evaluated with the help of the modified GBZ theory \cite{SM}.

\begin{figure}[t]
\includegraphics [width=8.5 cm]{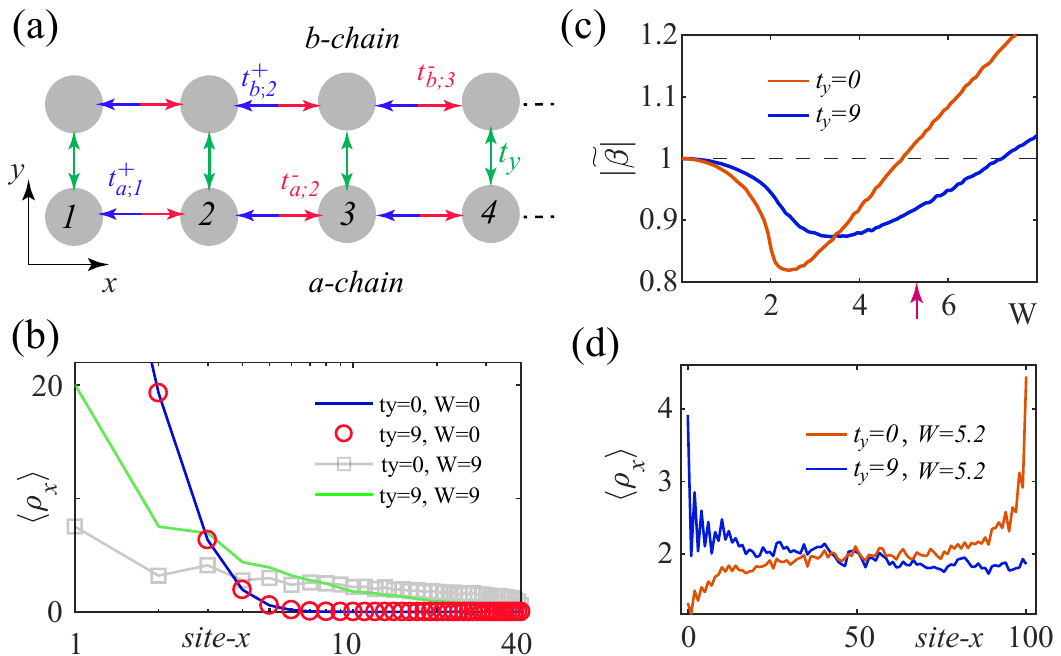}
\caption{(a) Schematic diagram for the double-chain model.
(b) $\langle\rho_x\rangle$ for one-chain ($t_y=0$) and double-chain ($t_y=9$) cases. The parameters are $t^+=4$, $t^-=1$ and $W=9$.
 (c) $|\widetilde{\beta}|$ versus $W$ for $t_y=0$ ($\widetilde{\beta}_l$) and $t_y=9$ ($\widetilde{\beta}_d$).
 (d) is the same with (b), except $t^+_{a,b;i}=1$, $t^-_{a,b;i}=1+w_{a,b;i}$.
 }
\label{f4}
\end{figure}

\textit{Coupling-reversed NHSE's directions.---} To verify the universality of our theory, we next study the $\mathcal{BBC}$ for multi-band cases. Taking the double-chain model in Fig. \ref{f4}(a) as an example, its Hamiltonian reads \cite{SM}:
\begin{equation}
\mathcal{H}_{d}=\sum_{i;k\in[a,b]}-t^\pm_{k;i}c^\dagger_{k;i}c_{k;i\pm1}
-t_yc^\dagger_{a;i}c_{b;i}-t_yc^\dagger_{b;i}c_{a;i}.
\end{equation}
 One requires $t^\pm_{a/b;i}=t^{\pm}+w^{\pm}_{a/b;i}$ and $w^{\pm}_{a/b;i}\in[-\frac{W}{2},\frac{W}{2}]$.
For $t_y=0$, one has $\widetilde{\beta}_{a/b}^{2N}=\prod_i [t_{a/b;i}^-/t_{a/b;i}^+]$ with $\widetilde{\beta}_{a}=\widetilde{\beta}_{b}$.
However, for $|t_y|>|t^\pm_{y;i}|$, the coefficient satisfies \cite{SM}:
\begin{equation}
\widetilde{\beta}_d^{2N}\approx \prod_{i=1}^{N} [\frac{2t^- + w_{a;i}^- + w_{b;i}^-}{2t^+ + w_{a;i}^+ + w_{b;i}^+}].
\label{betad}
\end{equation}
It is different from $\widetilde{\beta}_{a/b}$ when $W\neq0$.
Thus, the NHSEs show distinct features for $t_y=0$ and $|t_y|>|t^\pm_{a,b;i}|$ compared with the $W=0$ case, which is confirmed by the variation of $\langle\rho_x\rangle=\langle\sum_{k,i}|\psi_{k,i}(x)|^2\rangle$ in Fig. \ref{f4}(b).
More significantly, the differences also lead to the coupling-reversed NHSE's directions.
 By coupling two disordered Hatano-Nelson chains each with NHSE in the right directions [$\widetilde{\beta}_{a/b}>1$], the NHSE for the whole system is in the left direction instead [$\widetilde{\beta}_d<1$], shown in Figs. \ref{f4}(c)-(d) \cite{footnote7}. Such a feature can only be identified by our theory and has no Hermitian counterparts.
  We notice that the disordered Hatano-Nelson model has been experimentally realized \cite{disorder12}.
The coupling-reversed NHSE's directions gives unique approaches to manipulating the NHSEs, which is exclusive for disordered non-Hermitian systems.

\textit{Conclusion and Discussion.---}
In summary, we studied the $\mathcal{BBC}$ in the non-Hermitian system $\mathcal{H}$ with the combination of the disorder effect and NHSEs.
We demonstrate that the $\mathcal{BBC}$ for samples without translational symmetry requires: (i) A transformation $\mathcal{H}\rightarrow\mathcal{\widetilde{H}}$ that preserves the determinant under $\mathcal{OBC}$ as $\det[E-\mathcal{H_{OBC}}]=\det[E-\mathcal{\widetilde{H}_{OBC}}]$; (ii) The transformed Hamiltonian possesses the $\mathcal{BBC}$ as $\det[E-\mathcal{\widetilde{H}_{OBC}}]
\approx\det[E-\mathcal{\widetilde{H}_{PBC}}]$.
Based on these two requirements, we constructed the $\mathcal{BBC}$ in some typical disordered non-Hermitian systems and obtained the modified GBZ theory. The modified transformation parameter is the order parameter characterizing the NHSE, and the disorder effect on NHSE is uncovered.
Based on our analytical results, the disorder-enhanced and disorder-irrelevant NHSEs as well as coupling-reversed NHSE's directions are obtained. These features can be confirmed in experiments \cite{ref23,disorder12}.

The requirements in Eq. (\ref{EQ4}) to achieve $\mathcal{BBC}$ should be suitable for all the non-Hermitian systems.
In SM. V \cite{SM}, the generalization of our approach is clarified.
Details for samples with long-range hopping and higher dimensions is available in SM. VI \cite{SM}. The applicability of our approach for on-site disorder cases is verified in \cite{MGBZOS}.
Our work unveils the mechanism of $\mathcal{BBC}$ and extends the manipulation of NHSEs in disordered non-Hermitian systems.
By modulating the disorder schemes, one can manage to obtain the NHSE with required features in both theoretical and experimental aspects.

\textit{Acknowledgements.---}
We are grateful for fruitful discussions with Q. Wei, Z. S. Yang, M. Lu, Y. J. Wu and J. Yu.
This work was supported by the National Basic Research Program of China (Grant No. 2019YFA0308403), NSFC under Grant No. 11822407 and No. 12147126, and a Project Funded by the Priority Academic Program Development of Jiangsu Higher Education Institutions.

\begin{widetext}
\newpage
%\counterwithin{equation}{section} % reset the counting for the figure captions.

%\counterwithin{figure}{section} % reset the counting for the figure captions.

\setcounter{equation}{0}
\setcounter{figure}{0}
\setcounter{table}{0}
%\setcounter{page}{1}
%\makeatletter
\renewcommand{\theequation}{S\arabic{equation}}
\renewcommand{\thefigure}{S\arabic{figure}}
\renewcommand{\bibnumfmt}[1]{[S#1]}
\renewcommand{\citenumfont}[1]{S#1}

\begin{center}
\textbf{Supplementary Materials for ``Bulk-Bulk Correspondence in Disordered Non-Hermitian Systems"}
\end{center}

\begin{center}
Zhi-Qiang Zhang$^{1,2}$, Hongfang Liu$^{1,2}$, Haiwen Liu$^{3}$, Hua Jiang$^{1,2,*}$, and X. C. Xie$^{4,5}$
\end{center}

\begin{center}
$^1$~{\it School of Physical Science and Technology, Soochow University, Suzhou 215006, China}

$^2$~{\it Institute for Advanced Study, Soochow University, Suzhou 215006, China}

$^3$~{\it Center for Advanced Quantum Studies, Department of Physics, Beijing Normal University, Beijing 100875, People's Republic of China}

$^4$~{\it International Center for Quantum Materials, School of Physics, Peking University, Beijing 100871, China}

$^5$~{\it CAS Center for Excellence in Topological Quantum Computation, University of Chinese Academy of Sciences, Beijing 100190, China}

\end{center}

\tableofcontents
%%%%%%%%%%%%%%%%%%%%%%%%%%%%%

\section{ Introduction for the Supplementary Materials}

In this Supplementary Materials, we present more details of the bulk-bulk correspondence (BBC) proposed in the main text.
  {\it Specifically, the BBC is defined as: the consistency between the eigenvalues calculated under open and periodic boundary conditions.}
 In Section \B{II}, the details of the modified generalized Brillouin zone (GBZ) theory for disordered Hatano-Nelson models are exhibited.
Sections \B{III} gives additional numerical results.
  In Section \B{IV}, the relations between our theory and the winding number approach for the Hatano-Nelson model are clarified.
 Section \B{V} presents the generalization of our approach to different models.
 Applications of our theory for models with long-range hopping and {\it etc.} are available in Section \B{VI}.
In Section \B{VII}, some details of the derivations and proofs are available.

\section{ BBC and modified GBZ for disordered Hatano-Nelson model}

We present the {\it renormalized} $\beta$ for the modified GBZ theory of disordered Htano-Nelson model in this section. Based on the results in Sec. \B{VII. A} and Sec. \B{VII. B}, one has:
\begin{align}
\begin{split}
\widetilde{\beta}^{2n}=[\frac{t_Lt_1^-t_2^-\cdots t_{m}^-\cdots t_{n-1}^-}{t_Rt_1^+t_2^+\cdots t_{m}^+\cdots t_{n-1}^+}].
\end{split}
\end{align}

\subsection{A. Clean samples: revisit GBZ theory}

For clean samples one has:
\begin{align}
\begin{split}
&t^-_{m}\equiv t^-_1=t-\gamma;~~t^+_{m}\equiv t^+_1=t+\gamma.
\end{split}
\end{align}
Thus, one has $\beta^{2n}=\frac{(t_1^-)^n}{(t_1^+)^n}$ and $\beta=\sqrt{\frac{t^-_1}{t^+_1}}=\sqrt{\frac{t-\gamma}{t+\gamma}}$, which is the same as those in the GBZ theory.

\subsection{B. Dirty samples: modified GBZ theory and renormalized transformation parameter}

For a dirty sample considered, one has
\begin{align}
\begin{split}
\widetilde{\beta}^{2n}=[\frac{t_Lt_1^-t_2^-\cdots t_{m}^-\cdots t_{n-1}^-}{t_Rt_1^+t_2^+\cdots t_{m}^+\cdots t_{n-1}^+}].
\end{split}
\end{align}
Therefore, the renomalized $\widetilde{\beta}$ for a specific disordered sample considered in the main text [$t^+_{m}=t+\gamma+w^+_m$ and $t^-_{m}=t-\gamma+w^-_{m}$ ]
 can be marked as:
\begin{align}
\begin{split}
\widetilde{\beta}^{2n}=& \frac{\prod_{m=1}^{n} (t-\gamma+w^-_m)}{\prod_{m=1}^{n} (t+\gamma+w^+_m)}.
\end{split}
\end{align}
For $w_m^{\pm}\in[-\frac{W}{2},\frac{W}{2}]$,
the renormalized $\beta$ can be obtained as follows \cite{SSL6}:
\begin{align}
\begin{split}
\ln(\widetilde{\beta})&=\lim_{n\rightarrow\infty}\frac{1}{2n}[\sum_{m=1}^{n}\ln(t-\gamma+w^-_m)
-\sum_{m=1}^{n}\ln(t+\gamma+w^+_m)]\\
&=\int_{-W/2}^{W/2}\ln(t-\gamma+x)\frac{dx}{2W}
-\int_{-W/2}^{W/2}\ln(t+\gamma+x)\frac{dx}{2W}.
\end{split}
\end{align}
Finally, we capture the second key point: the analytical results for the renormalized $\beta$ versus disorder strength $W$ and $\gamma$
\begin{align}
\begin{split}
\widetilde{\beta}=[\frac{(t-\gamma+\frac{W}{2})^{t-\gamma+\frac{W}{2}}(t+\gamma-\frac{W}{2})^{t+\gamma-\frac{W}{2}}}
{(t-\gamma-\frac{W}{2})^{t-\gamma-\frac{W}{2}}(t+\gamma+\frac{W}{2})^{t+\gamma+\frac{W}{2}}}]^{\frac{1}{2W}}.
\label{EQS43}
\end{split}
\end{align}
The average of $\beta$ is similar to the self-average in disordered mesoscopic samples. More details of the deviation of Eq. (\ref{EQS43}) is available in Sec. \B{VII. E}. It is not an average on different samples. Thus, $\widetilde{\beta}$ is actually a well-defined and unified transformation parameter when $n$ is large enough.

\subsection{C. Modified GBZ theory for models in Fig. 3 in the main text}

 \begin{figure*}[h]
\includegraphics[width=16cm]{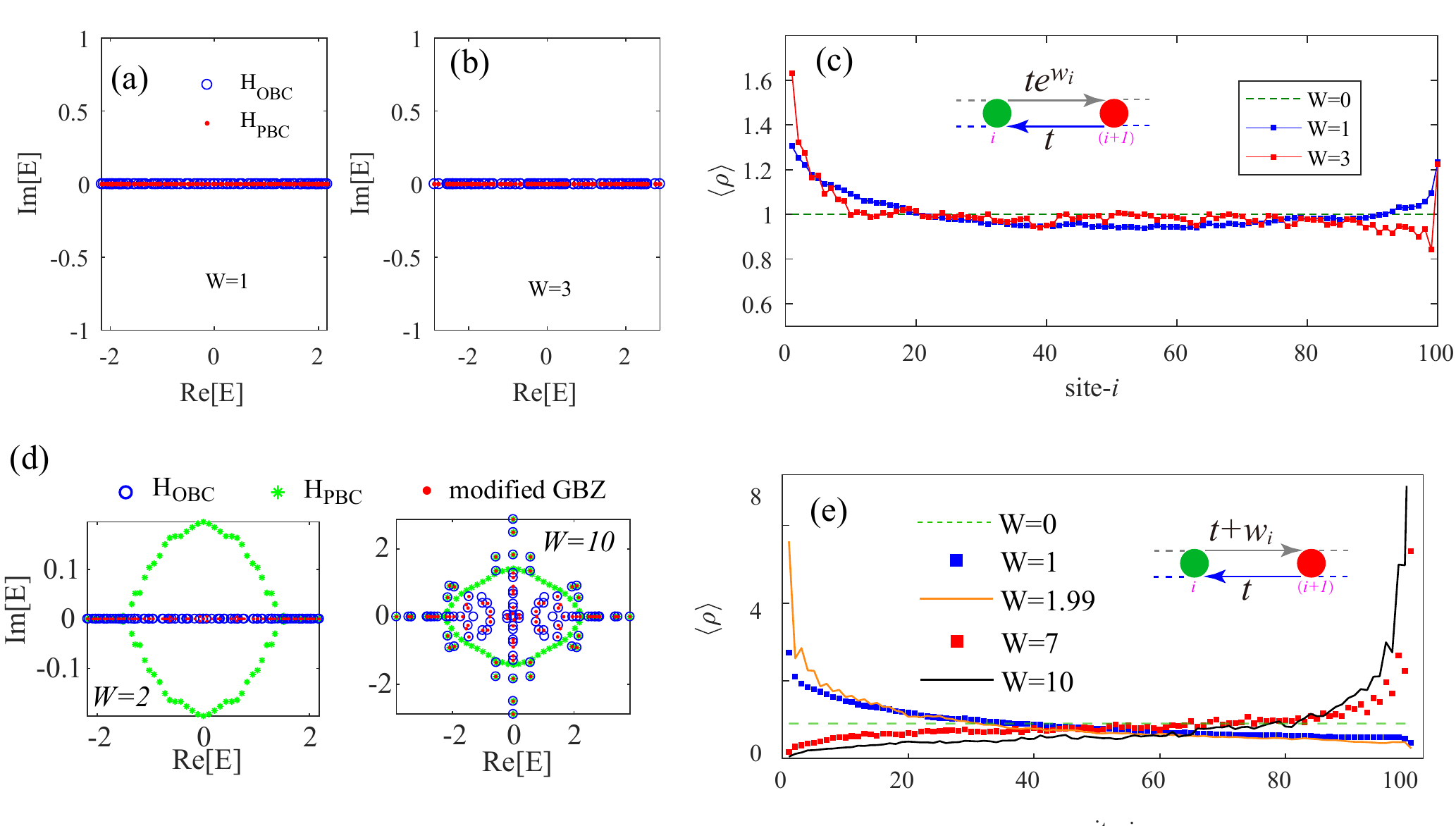}
\caption{(Color online). (a) and (b) $\mathrm{Re}[E]$ versus $\mathrm{Im}[E]$ for different disorder strengths. The Hamiltonian reads $\mathcal{H_R}= \sum_i-tc_{i}^\dagger c_{i+1}-te^{w_i}c_{i+1}^\dagger c_{i}$, where $t^+_i=t$ and $t^-_{i}=te^{w_i}$ with $w_i\in[-\frac{W}{2},\frac{W}{2}]$. $W$ is the disorder strength.
$H_{\mathrm{OBC}}$ and $H_{\mathrm{PBC}}$ correspond to the initial Hamiltonian under open and periodic boundary conditions, respectively.
(c) $\langle\rho\rangle$ for different disorder strengths. Except $t^+_i=t$ and $t^-_{i}=t+w_i$, (d) and (e) are the same with (b) and (c), respectively.
 }
\label{S0}
\end{figure*}

For model $\mathcal{H_R}= \sum_i-tc_{i}^\dagger c_{i+1}-te^{w_i}c^\dagger_{i+1}c_i$, one has $t^+_i=t$ and $t^-_{i}=te^{w_i}$ with $w_i\in[-\frac{W}{2},\frac{W}{2}]$.  We set $t=1$. Thus,
\begin{align}
\begin{split}
\widetilde{\beta}_r^{2n}=[\frac{t_Lt_1^-t_2^-\cdots t_{m}^-\cdots t_{n-1}^-}{t_Rt_1^+t_2^+\cdots t_{m}^+\cdots t_{n-1}^+}]=\frac{\prod_{m=1}^{n} te^{w_m}}{t^n}=\prod_{m=1}^{n} e^{w_m}.
\end{split}
\end{align}
One has:
\begin{align}
\begin{split}
\ln(\widetilde{\beta}_r)&=\lim_{n\rightarrow\infty}
\frac{1}{2n}[\sum_{m=1}^{n}\ln(e^{w_{m}})]=\frac{1}{2n}[\sum_{m=1}^{n}w_{m}]=0
\end{split}
\end{align}
Thus, $\widetilde{\beta}_r=1$ for different disorder strengths and the NHSE is absent.
As shown in Figs. \ref{S0}(a) and (b), eigenvalues calculated under $PBC$ show the absence of the close loop characteristics. Furthermore, the eigenvalues calculated under $OBC$ and $PBC$ spectrums fit perfectly. The density distributions $\langle\rho\rangle=\sum_{i\in \mathrm{all}}|\psi_i|^2$ in Fig. \ref{S0}(c) also verifies the fact that the NHSE feature is absent. The deviation comes from the finite size effect. These results are consistent with our analysis.

Similarly, for model $\mathcal{H_L}= \sum_i-tc_{i}^\dagger c_{i+1}-(t+w_{i})c^\dagger_{i+1}c_i$, one has $t^+_i=t$ and $t^-_{i}=t+w_{i}$ with $w_i\in[-\frac{W}{2},\frac{W}{2}]$.
One has
\begin{align}
\begin{split}
\widetilde{\beta}_l^{2n}=[\frac{t_Lt_1^-t_2^-\cdots t_{m}^-\cdots t_{n-1}^-}{t_Rt_1^+t_2^+\cdots t_{m}^+\cdots t_{n-1}^+}]=\frac{\prod_{m=1}^{n} (t+w_m)}{t^n}=\prod_{m=1}^{n} (1+w_m).
\end{split}
\end{align}
Thus,
\begin{align}
\begin{split}
\ln(\widetilde{\beta}_l)=&\lim_{n\rightarrow\infty}
\frac{1}{2n}[\sum_{m=1}^{n}\ln(1+w_m)]=\int_{-W/2}^{W/2}\ln(1+x)\frac{dx}{2W}
=\ln[(\frac{(1+\frac{W}{2})^{1+\frac{W}{2}}e^{-W}}{(1-\frac{W}{2})^{1-\frac{W}{2}}}
)^{\frac{1}{2W}}].
\end{split}
\end{align}
Finally, one has $\widetilde{\beta}_l=[\frac{(1+\frac{W}{2})^{1+\frac{W}{2}}e^{-W}}{(1-\frac{W}{2})^{1-\frac{W}{2}}}
]^{\frac{1}{2W}}$. The $BBC$ is available by considering $\widetilde{\beta}_l$ as shown in Fig. \ref{S0}(d). The variation of $\langle\rho\rangle$ identifies the existence of NHSEs when disorder is considered [see Fig. \ref{S0}(e)].

\subsection{D. Relationships between Eigenfunctions for samples with BBC}

$\mathcal{H}$ and $\mathcal{\widetilde{H}}$ are the original and the transformed Hamiltonian, respectively. Here, $\mathcal{\widetilde{H}}=\mathcal{S}^{-1}\mathcal{H}\mathcal{S}$ and $\mathcal{S}=\mathrm{diag}[\widetilde{\beta},\widetilde{\beta}^2,\widetilde{\beta}^3,\cdots, \widetilde{\beta}^n]$.
They satisfy the eigen-equations as:
  \begin{align}
\begin{split}
\mathcal{H}\psi_i=E_i\psi_i; ~\mathcal{\widetilde{H}}\widetilde{\psi_i}=E_i\widetilde{\psi_i}.
 \end{split}
\end{align}
 The BBC ensures $\widetilde{\psi_i}\approx S^{-1}\psi_i$. The detectable quantity is the absolute value of the eigenvectors $|\widetilde{\psi_i}|$ and $|\psi_i|$ in which the phase of $\widetilde{\beta}$ is irrelevant. For a fixed $\psi_i$, the variation of $\widetilde{\psi_i}$ will also alter the $\widetilde{\beta}$. More details see also Ref. \cite{SROSMGBZ}

\section{ Numerical results for disordered models}

In this section, we give some numerical results to support our analytical discussions.

\subsection{A. Stability of our approach for different conditions }

 \begin{figure*}[h]
\includegraphics[width=16cm]{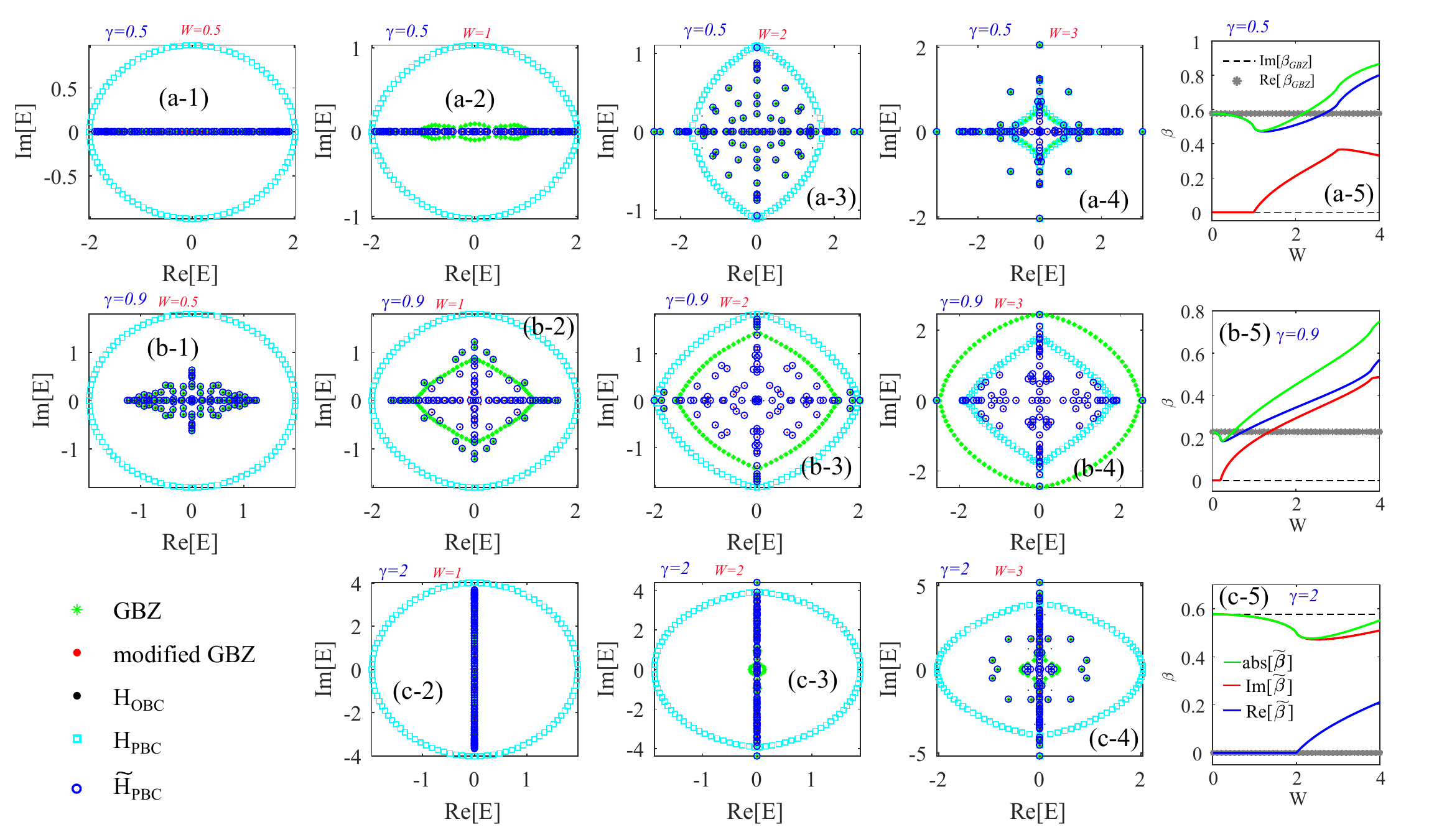}
\caption{(Color online). $\mathrm{Re}[E]$ versus $\mathrm{Im}[E]$ for different cases.
The disorder strength $W$ and $\gamma$ have been marked in the figure. Other parameters are $t=1$ and $N=100$. (a-5), (b-5) and (c-5) show the $\beta$ versus disorder strength.
GBZ corresponds to the cases with $\beta=\beta_{\mathcal{GBZ}}=\sqrt{\frac{t-\gamma}{t+\gamma}}$, and $t^+_i\rightarrow \beta t^+_i$, $t^-_{i}\rightarrow \beta^{-1} t^-_{i}$.
Modified GBZ is the same with GBZ except that $\beta_{\mathcal{GBZ}}$ is replaced by $\widetilde{\beta}$.
$H_{\mathrm{OBC}}$ and $H_{\mathrm{PBC}}$ correspond to the initial Hamiltonian under OBC and PBC, respectively.
$\widetilde{H}_{\mathrm{PBC}}$ corresponds to the Hamiltonian with $\widetilde{t}^+_i=\widetilde{t}^-_{i}= \sqrt{t^+_it_{i}^-}$ under the PBC.
 }
\label{S1}
\end{figure*}

Taking the Hatano-Nelson model as an example, the Hamiltonian reads \cite{SHN}:
\begin{align}
\begin{split}
\mathcal{H} =\sum_{i=1}^{N} -t^+_ic_{i}^\dagger c_{i+1}-t^-_{i}c_{i+1}^\dagger c_{i}.
\end{split}
\end{align}
with $t^+_i=(t+\gamma+w^+_i)$ and $t^-_{i}=(t-\gamma+w^-_{i})$.
The Anderson disorder is introduced as $w^{\pm}_{i}\in[-\frac{W}{2},\frac{W}{2}]$.
The stability of the transformation $\widetilde{t}^+_i=\widetilde{t}^-_{i}=\sqrt{t^+_it^-_{i}}$ with $\beta_i=\sqrt{\frac{t_i^-}{t_i^+}}$ and the modified GBZ [$\widetilde{\beta}$] are shown in Fig. \ref{S1}.

 The plot of GBZ corresponds to the results of the transformed Hamiltonian under $PBC$ with a transformation $t^+_i\rightarrow \beta_{\mathcal{GBZ}} t^+_i$ and $t^-_{i}\rightarrow \beta_{\mathcal{GBZ}}^{-1} t^-_{i}$, in which $\beta_{\mathcal{GBZ}}=\sqrt{\frac{t-\gamma}{t+\gamma}}$.
The modified GBZ is the same with the previous GBZ except that $\beta_{\mathcal{GBZ}}$ is replaced by $ \widetilde{\beta}$.
$H_{\mathrm{OBC}}$ and $H_{\mathrm{PBC}}$ correspond to results calculated by using the original Hamiltonian under OBC and PBC, respectively.
$\widetilde{H}_{\mathrm{PBC}}$ corresponds to the effective Hamiltonian with $\widetilde{t}^+_i=\widetilde{t}^-_{i}= \sqrt{t^+_it^-_{i}}$ under the PBC.

 As one can see, results for the modified GBZ and $\widetilde{H}_{\mathrm{PBC}}$ overlap with $H$ under the OBC for different parameters. In addition, one should notice that the results for GBZ with $\beta_{\mathcal{GBZ}}=\sqrt{(t-\gamma)/(t+\gamma)}$ is the same with the modified GBZ as well as the $H_{\mathrm{OBC}}$ for weak disorder since $|\beta_{\mathcal{GBZ}}|\approx |\widetilde{\beta}|$ as shown in Figs. \ref{S1}(a-5),(b-5) and (c-5). Nevertheless, GBZ is totally wrong when disorder is strong enough.

 Taking $\gamma=0.5$ as an example, GBZ give the correct result as shown in Fig. \ref{S1}(a-1), since $\widetilde{\beta}\approx\beta_{\mathcal{GBZ}}\approx0.577$ [see Fig. \ref{S1}(a-5)]. By increasing $W$, the results of GBZ deviate from the open boundary results, as shown in Fig. \ref{S1} (a-2) since $\widetilde{\beta}$ deviates from $\beta_{\mathcal{GBZ}}\approx0.577$. By further increasing $W$, the result of GBZ overlap with the open boundary cases again [see Fig. \ref{S1} (a-3)] since $|\widetilde{\beta}|$ approaches to $0.577$ again. For stronger disorder strength, the GBZ transformation $\beta_{\mathcal{GBZ}}$ can not give correct results since $|\widetilde{\beta}|$ tends to approach one.

 These numerical results are consistent with our discussions and analytical results.

\subsection{B. Hopping disorder with exponential forms}

Based on the universal criterion of $\mathcal{BBC}$, one is able to qualitatively illuminate the universality of the disorder-dependent of $\beta_{\mathcal{GBZ}}$ and the eigenvalues for different types of disorder. For the nearest hopping model, (i) If the disorder preserves $t^+_it^-_{i}$, then the open boundary spectrum $E_{\mathcal{OBC}}$ is invariant since $\det[E-\mathcal{H_{OBC}}]= f(t_i^+t^-_{i},E)$;
 (ii) If the disorder preserves $\prod_it^-_{i}/t^+_{i}$, then the GBZ theory and $\widetilde{\beta}=\beta_{\mathcal{GBZ}}$ hold since $\widetilde{\beta}=\beta_{\mathcal{GBZ}}\propto\prod_i\frac{t^-_{i}}{t^+_i}$;
 (iii) If the disorder preserves $\prod_it^+_it^-_{i}$, $\prod_i t^+_i$ and $\prod_i t^-_{i}$, then $\det[E-\mathcal{H_{PBC}}]$ as well as $E_{\mathcal{PBC}}$ are invariant.
 If disorder does not preserve all these three classes of quantities, then nothing remains invariant.
For illustration, several types of disorder are considered, and the corresponding results are summarized in TABLE. \ref{TABLES1}.

\begin{table}[h]
\centering
\caption{ The influence of disorder on Hamiltonian
$\mathcal{H}_{1} = -\sum_i[(t+\gamma)e^{w^+_i}c_{i}^\dagger c_{i+1}+(t-\gamma)e^{w^-_{i}}c_{i+1}^\dagger c_{i}]$.
One has $t^+_i=(t+\gamma)e^{w^+_i}$ and $t^-_{i}=(t-\gamma)e^{w^-_{i}}$ for $\mathcal{H}_1$. $w^\pm_i$ is random numbers. $[w^\pm_i]$ stands for the sets of all possible values.
 $\widetilde{\beta}$ is the transformation coefficient of the modified GBZ theory. $E_\mathcal{PBC/OBC}$ is the eigenvalues calculated under different boundary conditions.
{\it Fixed/changed} suggests the value is unchanged/chagned compared with disordered cases. {\it Depends} means that it depends.
 }\label{TAB1}
\setlength{\tabcolsep}{6mm}{
\begin{tabular}{lccccc}
\hline
\hline
disorder type &  $w^\pm_i\in$  & $\widetilde{\beta}$  & $E_\mathcal{PBC}$ & $E_\mathcal{OBC}$ & example\\
\hline
$w^-_i=-w^+_{i}$ & any               &   depends     &  depends     & fixed & Fig. \ref{S2}(c)/(d)\\
$w^-_i=-w^+_{i}$ & $[\frac{-W}{2},\frac{W}{2}]$ & fixed  & fixed  & fixed & Fig. \ref{S2}(c)\\
$[w^-_i]=[w^+_{i}]$ & any                & fixed  &   changed    & changed & Fig. \ref{S2}(b) \\
%$\varepsilon_m$ & $\mathbb{R}$  & fixed   & changed   & changed & Fig. \ref{S3}(a)-(d)\\
\hline
\hline
\end{tabular}}
\label{TABLES1}
\end{table}

 \begin{figure*}[h]
\includegraphics[width=15cm]{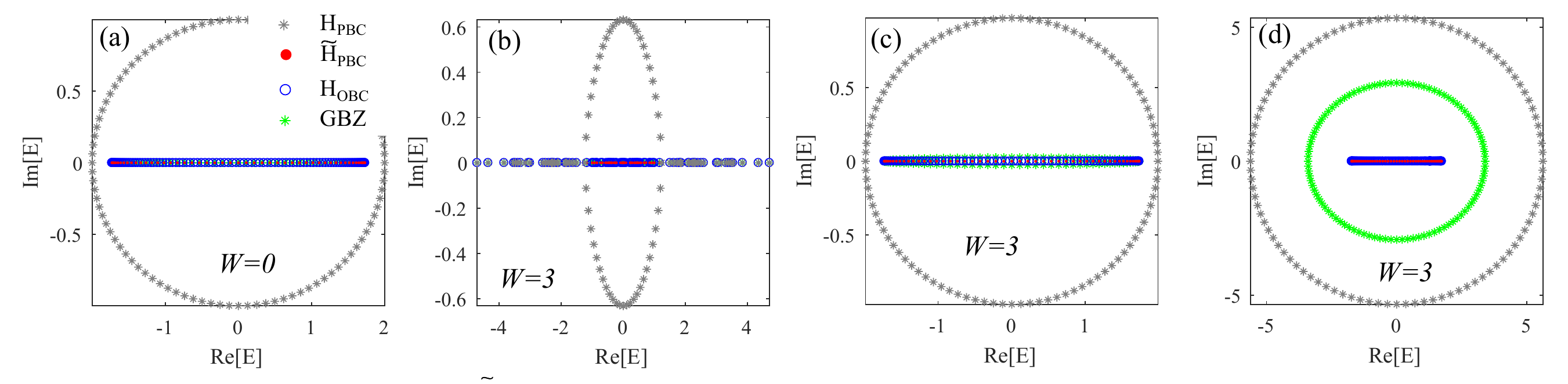}
\caption{(Color online). $\mathrm{Re}[E]$ versus $\mathrm{Im}[E]$.
The disorder strength $W$ has been marked in the figure. Other parameters are $t=1$, $\gamma=0.5$ and $n=100$. (a) without disorder. (b) $w^-_i=w^+_{i}$; (c) $w^-_i=-w^+_{i}$ with $w^\pm_i\in[-\frac{W}{2},\frac{W}{2}]$. (d) $w^-_i=-w^+_{i}$ with $w^+_i\in[0,W]$. $W$ is the disorder strength. }
\label{S2}
\end{figure*}

We check the stability of the results shown in TABLE. \ref{TABLES1}. The Hamiltonian reads:
\begin{align}
\begin{split}
\mathcal{H} =\sum_{i=1}^{n} -t^+_ic_{i}^\dagger c_{i+1}-t^-_{i}c_{i+1}^\dagger c_{i}.
\end{split}
\end{align}
where the hopping terms are set as $t_i^+=(t+\gamma)e^{w^+_i}$ and $t_i^-=(t-\gamma)e^{w^-_i}$.
The three different kinds of disorder calculated are:

(i). For $w^+_{i}=-w^-_i$, then $t^+_it^-_{i}$ is an invariant. Based on Eq. (\ref{EQS23}), the open boundary spectrum remains unchanged no matter which kind of disorder $w^{\pm}_i$ is applied [see Figs. \ref{S2} (a), (c) and (d)].

(ii). If $w^+_i=-w^-_{i}$ and $w^{\pm}_{i}\in [-\frac{W}{2},\frac{W}{2}]$, then $\det [E-\mathcal{H}_{\mathcal{PBC}}]$ and $\beta_{\mathcal{GBZ}}$ are invariant based on Eqs. (\ref{EQS26}) and (\ref{EQSd}).
 Thus, the periodic boundary spectrum also remains unchanged, no matter how large $W$ is [as shown in Fig. \ref{S2} (a) and (c)]. Generally, $\beta_{\mathcal{GBZ}}$ is invariant when the sample size $n$ is large enough with $\prod_{i=1}^n e^{w^{\pm}_{i}}\sim1$. Furthermore, $\det [E-\mathcal{H}_{\mathcal{PBC}}]$ remains almost unchanged since $\prod_{i=1}^n t^+_{i}\sim (t+\gamma)^n$, $\prod_{i=1}^n t^-_{i}\sim (t-\gamma)^n$ and $t^+_it^-_{i}=t^2-\gamma^2$ holds.

(iii). If $w^+_i=w^-_{i}$, then $\beta_{\mathcal{GBZ}}$ is invariant based on Eq. (\ref{EQSd}). As shown in Fig. \ref{S2}(b), the results of GBZ still overlap with the open boundary cases when $W\neq0$.

\subsection{C. BBC for Su-Schrieffer-Heeger model}

In this section, the Su-Schrieffer-Heeger model \cite{SGBZ1} is considered
\begin{align}
\begin{split}
\mathcal{H}_{\mathrm{SSH}} =\sum_i-[t_0c_{i}^\dagger c_{i+1}+h.c.]-
 [(t+\gamma+w^+_{i+1})c_{i+1}^\dagger c_{i+2}+(t-\gamma+w^-_{i+1})c_{i+2}^\dagger c_{i+1}].
\label{EQSSH}
\end{split}
\end{align}
with $i\in \left\{1,3,5,\cdots,2m-1\cdots\right\}$. We mark the hopping term as follows
 \begin{align}
\begin{split}
 & t_{i+1}^+=(t+\gamma+w^+_{i+1}),\\
 & t_{i+1}^-=(t-\gamma+w^-_{i+1}),\\
 & t_{i}^{\pm}=t_0.
 \end{split}
\end{align}

We introduce the transformation $t^{\pm}_{i}\equiv t_0\rightarrow t_0$, $t^+_{i+1} \rightarrow\beta t^+_{i+1}$ and $t^-_{i+1}\rightarrow \beta^{-1}t^-_{i+1}$ for the clean samples.
One has:
\begin{align}
\begin{split}
\beta^{n}=[\frac{t_Lt_1^-t_2^-\cdots t_{m}^-\cdots t_{n-1}^-}{t_Rt_1^+t_2^+\cdots t_{m}^+\cdots t_{n-1}^+}].
\end{split}
\end{align}
Thus, $\beta=\sqrt{\frac{t_2^-}{t_2^+}}=\sqrt{\frac{t-\gamma}{t+\gamma}}$. Such a result is consistent with the previous studies \cite{SGBZ1,SGBZ2}.

When disorder is introduced, the transformation follows a unified parameter $\widetilde{\beta}$ that:
$t^+_{i+1} \rightarrow\widetilde{\beta} t^+_{i+1}$ and $t^-_{i+1}\rightarrow \widetilde{\beta}^{-1}t^-_{i+1}$ with $i\in\left\{1,3,5\cdots\right\}$.
Then the elimination of the asymmetric hopping of $\mathcal{\widetilde{H}}$ requires:
 \begin{align}
\begin{split}
\widetilde{\beta}^{n/2}[t_Rt^+_1t^+_2\cdots t_{m}^+\cdots t_{n-1}^+ ]=(\widetilde{\beta})^{-n/2}[t_Lt_1^-t_2^-\cdots t_{m}^-\cdots t_{n-1}^- ].
 \end{split}
\end{align}
which is equivalent to
  \begin{align}
\begin{split}
(\widetilde{\beta})^{n}=\frac{t_Lt_1^-t_2^-\cdots t_{m}^-\cdots t_{n-1}^-}{t_Rt^+_1t^+_2\cdots t_{m}^+\cdots t_{n-1}^+ }
=\frac{\prod_{m=1}^{n/2} (t-\gamma+w^-_m)}{\prod_{m=1}^{n/2} (t+\gamma+w^+_m)}.
\label{EQS52}
 \end{split}
\end{align}
Thus, the $\widetilde{\beta}$ here is the same as that in the Hatano-Nelson model.

 \begin{figure*}[h]
\includegraphics[width=15cm]{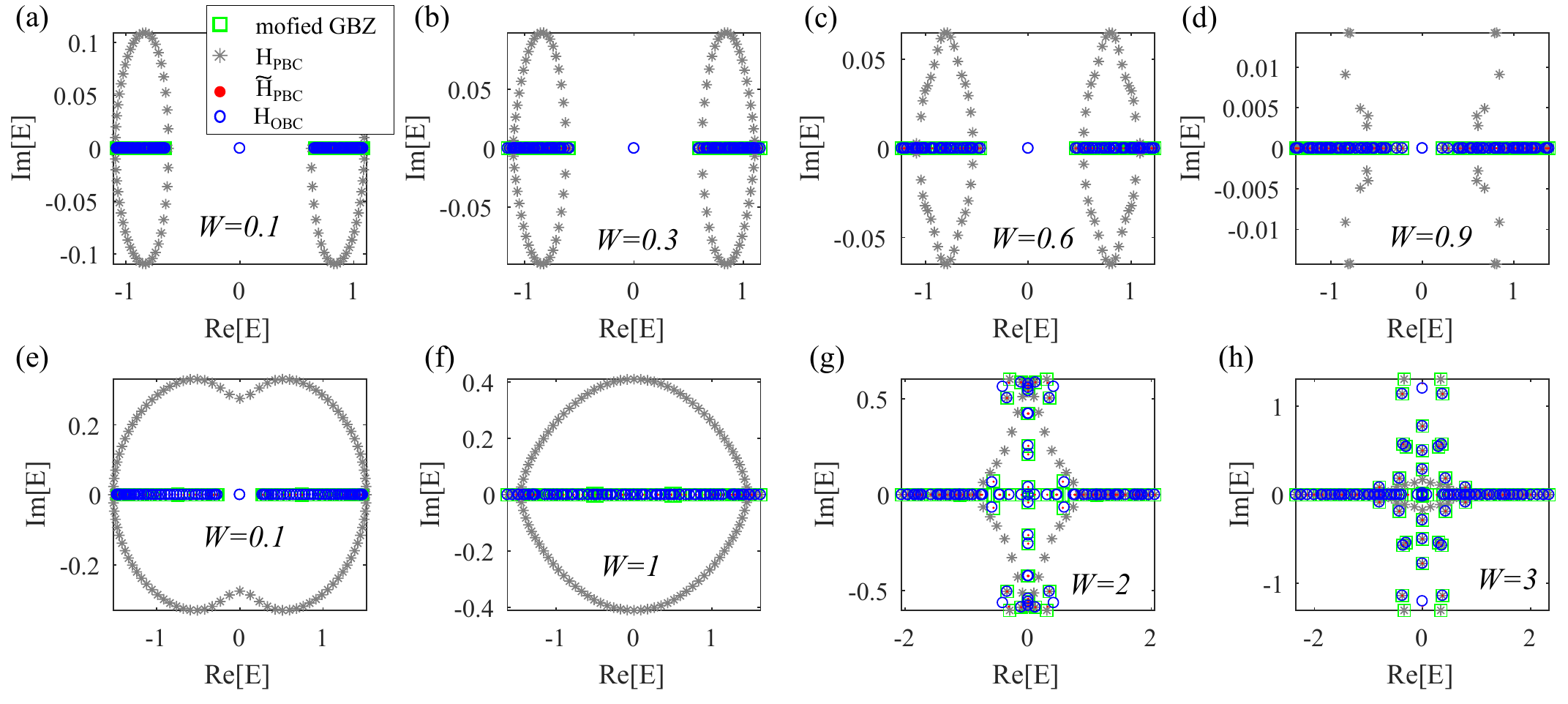}
\caption{(Color online). $\mathrm{Re}[E]$ versus $\mathrm{Im}[E]$.
The disorder strength $W$ has been marked in the figure. Other parameters are $t=1$, $\gamma=0.5$ and $N=50$. (a)-(d) $t_0=0.2$; (e)-(h) $t_0=0.6$.  }
\label{S4}
\end{figure*}

We pay attention to the model in Eq. (\ref{EQSSH}).
In Eq. (\ref{EQS52}), we have pointed out that the Su-Schrieffer-Heeger model has the same modified GBZ theory and the transformation parameter $\widetilde{\beta}$ as those in the Hatano-Nelson model. $\widetilde{\beta}$ is only applied to the translational symmetry. $t_0$ is fixed at its initial value.
The modified GBZ with transformation parameter $\widetilde{\beta}$ is the same as those in Eq. (\ref{EQS43}).

As clearly shown in Fig. \ref{S4}, the modified GBZ theory overlaps perfectly with the bulk spectrum under OBC. Notably, the zero eigenvalues are only available for the open boundary cases, which shows the periodic boundary features for other cases. Thus, our result is applicable for the Su-Schrieffer-Heeger model. These results strongly support our theory.

\section{ Relations between our theory and the winding number for the Hatano-Nelson model}

We distinguish the similarities and differences between our theory and the winding number approach.
 We start from the following formula:
\begin{align}
\begin{split}
 \det[E-\mathcal{H}_{\mathcal{PBC};~n\times n}]=&\det[E-\mathcal{H}_{\mathcal{OBC};~n\times n}]+f_{\mathcal{PBC}},
\label{REQ1}
\end{split}
\end{align}
with
$
f_{\mathcal{PBC}}=(-1)^{\tau}t_Lt_R\det[E-\mathcal{H}_{\mathcal{OBC};~n-2\times n-2}]+(-1)^{n+1}[\prod\limits_{i=1}^{n-1} t_Rt^+_i+\prod\limits_{i=1}^{n-1}t_Lt^-_{i}].
$
To calculate the winding number, one introduces a generalized phase with $t_R\rightarrow t_Re^{j\psi}$ and $t_L\rightarrow t_Le^{-j\psi}$.
Then, the winding number for the disordered samples can be calculated as follows \cite{NHS10,SdisTaler}:
\begin{align}
\begin{split}
\mathcal{N}=\frac{1}{2\pi j}\int_{0}^{2\pi}\frac{\partial\{\log\{\det[E-\mathcal{H}_{PBC}(\psi)]\}\}}{\partial{\psi}}d\psi
\end{split}
\end{align}
For fixed $E$, Eq. (\ref{REQ1}) can be rewritten as:
\begin{align}
\begin{split}
 \det[E-\mathcal{H}_{\mathcal{PBC};~n\times n}]=&(-1)^{n+1}[\prod\limits_{i=1}^{n-1} t_Rt^+_ie^{j\psi}+\prod\limits_{i=1}^{n-1}t_Lt^-_{i}e^{-j\psi}]+\kappa\\
 =&(-1)^{n+1}[T_1e^{j\psi}+T_2e^{-j\psi}]+\kappa.
\end{split}
\end{align}
We set $T_1=x_1^n=\prod\limits_{i=1}^{n-1} t_Rt^+_i$,
 $T_2=x_2^n=\prod\limits_{i=1}^{n-1}t_Lt^-_{i}$ and $\kappa=\det[E-\mathcal{H}_{\mathcal{OBC};~n\times n}]+(-1)^{\tau}t_Lt_R\det[E-\mathcal{H}_{\mathcal{OBC};~n-2\times n-2}]$.
For samples with NHSEs, one has $|T_1|\neq |T_2|$. If $n\rightarrow\infty$, it is reasonable to drop $T_2$ ($T_1$) after considering $x_1>x_2$ ($x_1<x_2$). Thus, one has:
 \begin{align}
\begin{split}
 \det[E-\mathcal{H}_{\mathcal{PBC};~n\times n}]
 =&(-1)^{n+1}T_{1,2}e^{\pm j\psi}+\kappa.
\label{REQ2}
\end{split}
\end{align}
The winding number is obtained as following:
\begin{align}
\begin{split}
\mathcal{N}=\frac{1}{2\pi
}\int_{0}^{2\pi}\frac{\pm(-1)^{n+1} T_{1,2}e^{\pm j\psi}}{\kappa+(-1)^{n+1}T_{1,2}e^{\pm j\psi}}d\psi
\label{inter1}
\end{split}
\end{align}
where
$
\kappa=\det[E-\mathcal{H}_{\mathcal{OBC};~n\times n}]+(-1)^{\tau}t_Lt_R\det[E-\mathcal{H}_{\mathcal{OBC};~n-2\times n-2}].
$

If $E$ approaches the eigenvalue of $\mathcal{H}_{\mathcal{OBC};~n\times n}$
[$\det[E-\mathcal{H}_{\mathcal{OBC};~n\times n}]\approx0$], the constance $\kappa$ could be ignored. Thus, one has $\det[E-\mathcal{H}_{\mathcal{PBC};~n\times n}]
 \approx (-1)^{n+1}T_{1,2}e^{\pm j\psi}$. By integrating Eq. (\ref{inter1}), the winding number with $\mathcal{N}=\pm1$ is obtained.

For samples without NHSEs, $|T_1|=|T_2|$ is guaranteed based on our minimization requirements. Taking $T_1=T_2$ as an example, one has
\begin{align}
\begin{split}
\mathcal{N}=\frac{1}{2\pi
j}\int_{0}^{2\pi}\frac{(-1)^{n+2}2T_1\sin\psi}{\kappa+(-1)^{n+1}2T_1\cos(\psi)}d\psi
\end{split}
\end{align}
When the $\kappa$ can be ignored, one has:
\begin{align}
\begin{split}
\mathcal{N}\approx\frac{1}{2\pi
j}\int_{0}^{2\pi}\frac{-2T_1\sin\psi}{2T_1\cos(\psi)}d\psi=0
\end{split}
\end{align}
If $\kappa$ is large enough ($E$ strongly deviate from the eigenvalue of $\mathcal{H}_{\mathcal{OBC};~n\times n}$), one also has $\mathcal{N}\propto \frac{1}{\kappa}=0$. Actually, $\frac{2T_1\sin\psi}{\kappa+(-1)^{n+1}2T_1\cos(\psi)}$ is an odd function, which ensures $\mathcal{N}=0$.

These results are consistent with our theory that $\beta=(T_2/T_1)^{1/2n}\neq1$ indicating the existence of NHSE. For $\beta<1$, one has $T_1>T_2$ with winding number $\mathcal{N}=1$. For samples without NHSEs, one has $\beta=1$ with $T_1=T_2$. In Fig. \ref{Re3}, we plot $\det[E-\mathcal{H}(\psi)]$ for $E=0$. Figure (a) shows the angle of $\det[\mathcal{H}(\psi)]$, and a non-trivial winding number exists for such a sample. Figures (b)-(d) shows that $\det[\mathcal{H}(\psi)]\propto (t^{-})^Ne^{-j\psi}=2^{40}e^{-j\psi}$, consisting with our discussions.

\begin{figure}[h]
   \centering
    \includegraphics[width=0.9\textwidth]{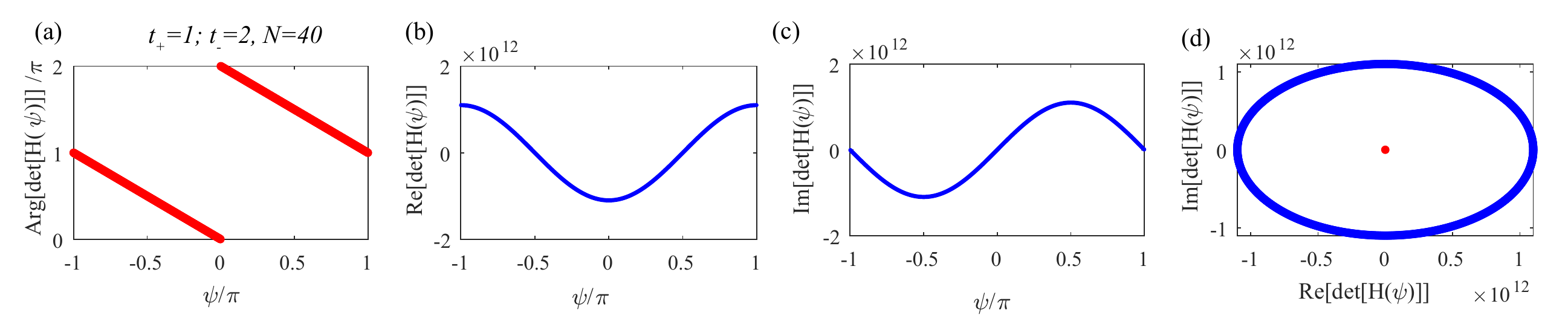}
    \caption{(Color online).(a) The evolution of $Arg[\det(\mathcal{H}(\psi))]$ versus the phase $\psi$ for non-Hermitian systems. The real and image parts are given in (b) and (c). (d) is the plot of $Re[det(\mathcal{H}(\psi))]$ versus $Im[det(\mathcal{H}(\psi))]$. The parameters have been marked in the figures. Following previous studies, we set $\psi\in[-\pi,\pi]$ in our numerical calculations.   }
   \label{Re3}
\end{figure}

\section{ General approach for BBC and its comparison with other approaches}

Our scheme for determining $\beta$ is highly general for almost all the non-Hermitian cases. Here, we propose the universal approach with only three steps as follows:\\

\fbox{\parbox{0.9\textwidth}{
 \B{\textbf{Step-1:}} One has to calculate the eigenvalues of $\mathcal{H}_{OBC}$. Specifically, if $E$ is not the eigenvalue of $\mathcal{H}_{OBC}$, $\det[E-\mathcal{H}_{OBC}]$ will give a vary large number, and the obtained $\beta$ could be meaningless.\\\\
 \B{\textbf{Step-2:}} One numerically or analytically calculates
\begin{equation}
\mathcal{F}(E,\beta)=|\det[E-\mathcal{H}_{OBC}]-\det[E-\mathcal{\widetilde{H}}_{PBC}(\beta)]|.
\end{equation}
For samples with size $N$, one requires
\begin{align}
\begin{split}
&\mathcal{\widetilde{H}_{OBC}(\beta)}=U\mathcal{H}_{OBC}U^{-1}
\end{split}
\end{align}
 with $U$ the diagonal matrix, i.e. $U=diag[\beta,\beta^2,\beta^3,\cdots]$. The periodic boundary condition is determined by the formula of $\mathcal{\widetilde{H}_{OBC}}$, and one obtains $\mathcal{\widetilde{H}_{PBC}(\beta)}$. A general choice of $\beta$ is clarified in the following.
 \\\\
 \B{\textbf{Step-3:}} For different $E$, the coefficient $\beta=\beta_{min}$ is available by minimizing
 $\mathcal{F}(E,\beta)$. The E-dependent of $\beta$ can be eliminated when a unified $|\beta|$ exists.
}}\\

Now, we clarify the universality of our approach in details. For $n$-$chain$ cases, the general approach should consider $\beta\in[\beta_1,\cdots, \beta_n]$ for each chain of a sample. However, by considering additional restrictions for the model, some redundant degrees of freedom [$\beta_i$] can be eliminated. To sum up, the multi-$\beta\in [\beta_1,\cdots]$ is the general condition. A typical example is the critical NHSEs \cite{NHSEs13} shown in Sec. VI. A.
 After considering additional restrictions of the model, some redundant degrees of freedom [$\beta_i$] can be eliminated. Specifically, one has $\beta_i=\beta_k$ while {\it``exchanging $i$-th and $k$-th chains with everything being almost unchanged"}.

\begin{table}[h]
\centering
\caption{ The comparisons of GBZ schemes, winding number approaches and our approach. $\varphi_n$ and $\varphi_1$ are the wavefunctions for different sites.
 }\label{TAB1}
\setlength{\tabcolsep}{1.1mm}{
\begin{tabular}{c|c|c|c}
\hline
\hline
\diagbox{Considerations}{Methods} & GBZ scheme & Winding number approach & Our approach\\[2ex]
\hline
Key equation & $\varphi_n=\beta^n\varphi_1$  & $\mathcal{N}=\oint\frac{\partial\{\ln\{\det[E-\mathcal{H_{PBC}}(\psi)]\}\}}{2\pi j\times\partial{\psi}}d\psi$  &
\makecell[r]{$\mathcal{F}=|\det[E-\mathcal{\widetilde{H}_{PBC}}(\beta)]$ \\ $-\det[E-\mathcal{H_{OBC}}]|$}\\[1ex]
\hline
translational symmetry & required & not required & \B{\textbf{not required}}\\[1ex]
\hline
characterization of NHSEs &$\beta e^{i\theta}$ & existence and directions & \makecell[c]{\B{\textbf{$\beta$: existence, directions}} \\ and \B{\textbf{strength}} }\\[1ex]
\hline
application limitations & clean samples only&  not suitable for critical NHSEs& \B{\textbf{highest universality}}\\[1ex]
\hline
dimension's requirements & unlimited& unlimited & \B{\textbf{unlimited}}\\[1ex]
\hline
\makecell[c]{ability of designing the \\  NHSEs' features} & clean samples only & no & \makecell[c]{\B{\textbf{both clean and}} \\ \B{\textbf{dirty samples}} }\\[1ex]
\hline
\makecell[c]{analytical formula for \\ dirty samples} &  \diagbox{}{} & unreported & \B{\textbf{this work}}\\[1ex]
\hline
\makecell[c]{limited to disorder schemes} & yes  & no & \B{\textbf{no}}\\[1ex]
\hline
\hline
\end{tabular}}
\label{TABLES2}
\end{table}

For clarity, the comparisons of GBZ schemes, winding number approaches and our approach are presented in TAB. \ref{TABLES2}.
 It clearly unveil that our approach combines the advantages of both GBZ theory and winding number approach and complements their shortages. Significantly, our approach is highly universal compared with other approaches.
 The universality of our theory also leads to the discovery of various new and unique features for non-Hermitian systems.
 In general, our approach is valid to systematic study the disorder effect in non-Hermitian systems, and its results also goes beyond the anticipations of the conventional GBZ theory.

\begin{figure}[t]
   \centering
    \includegraphics[width=0.8\textwidth]{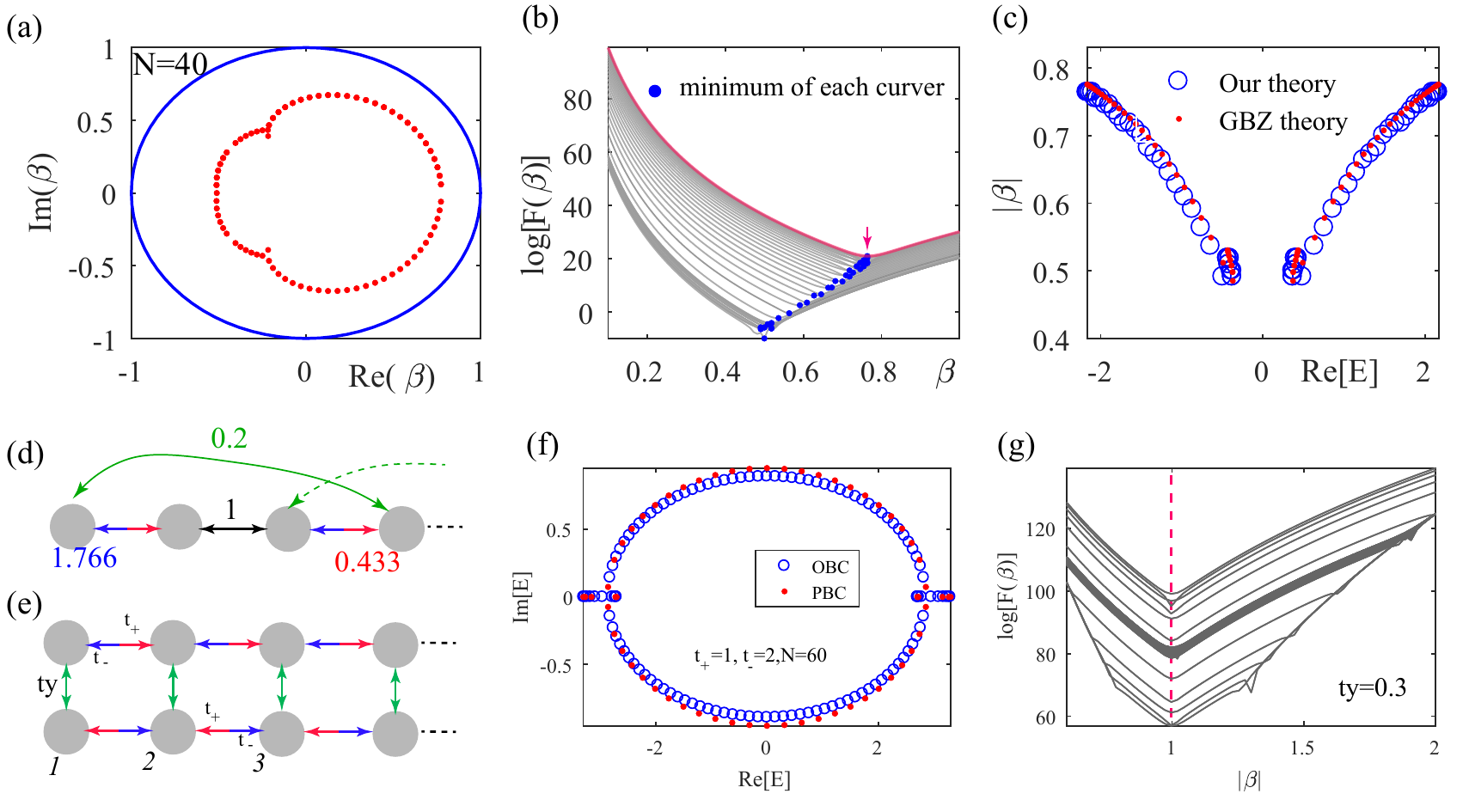}
    \caption{(Color online). (a)-(c) is corresponding to samples with long-range hopping shown in figure (d). (a) The plot of the generalized-Brillouin-zone (GBZ) by adopting the results in Ref. \cite{SGBZ2}. The parameters have been marked in the figure. (b) The plot of $\mathcal{F}(\beta)=|\det[E-\mathcal{H}_{OBC}]-\det[E-\mathcal{\widetilde{H}}_{PBC}(\beta)]|$ versus $\beta$ for different eigenvalues. The minimum of each curve has been marked by blue dots. (c) The plot of $|\beta|$ versus the real part of the eigenvalues $Re[E]$ for the $GBZ$ [red dots; data in figure (a)] and our scheme (blue circles; data in figure (b)). (f) shows the plot of $Re[E]$ versus $Im[E]$ for samples in (e). The evolution of $\mathcal{F}(\beta)$ is given in (g). The parameters have been marked in the figures. }
   \label{Re1}
\end{figure}

\section{ Applications of our theory beyond Hatano-Nelson model}

\begin{figure}[h]
   \centering
    \includegraphics[width=0.95\textwidth]{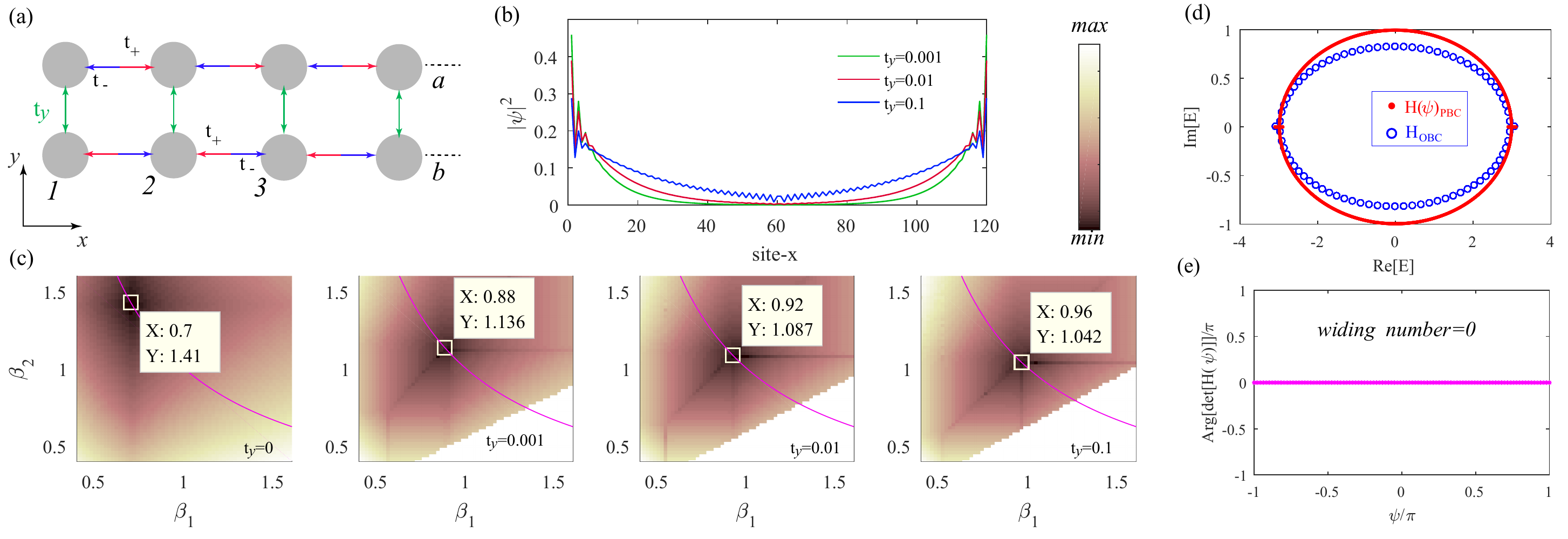}
    \caption{(Color online). (a) shows the model. (b) The plot of the eigenvectors with different $t_y$ for model in (a). We set sample size $N=60$. The eigenfunctions $|\psi|^2$ for the $a$ and $b$ chain are labeled by $site$-$x\in[1,60]$ and $site$-$x\in[61,120]$, respectively.  (c) shows the double parameters optimize of $\mathcal{F}(\beta_1,\beta_2)$. The white squares shows the minimum of $\mathcal{F}(\beta_1,\beta_2)$. The red solid lines are the plots of $\beta_1=1/\beta_2$.
    The parameters are $t_+=1$ and $t_-=2$.
    (d) plots the eigenvalues.
To determine the winding number, we adopt the standard approach with an additional flux $e^{j\psi}$ added on the periodic boundary hopping.
 $H(\psi)_{PBC}$ represent the eigenvalues for $H(\psi)$ under PBC with $\psi\in[-\pi,\pi]$, which forms a closed loop. $H_{OBC}$ shows the eigenvalues under OBC. The inconsistency between eigenvalues for $H_{OBC}$ and $H(\psi)_{PBC}$ verifies the crash of BBC and indicates the existence of NHSEs.
(e) The plot of $Arg[det[H(\psi)]]$ shows the absence of non-trivial winding number, which implies the absence of NHSEs.
 The sample size is fixed at $N=60$.  }
   \label{Re2}
\end{figure}

\subsection{A. Some typical examples: samples with long-rang hopping and samples with the critical NHSEs}
In general, our scheme is not limited to the Hatano-Nelson model and is suitable for more complicated models. Based on our proposed scheme, some interesting models with unique features are available.
In the following, we take some typical models as examples.

We take two typical models as examples. For models with long range hopping, as shown in Fig.\ref{Re1} (a), $\beta$ is energy $E$ dependent and the unified $|\beta|$ is not available \cite{SGBZ2}. Nevertheless, the minimum of $\mathcal{F}(\beta)$ for every eigenvalue precisely overlaps with the results of the GBZ curve \cite{SGBZ2}, shown in Fig.\ref{Re1} (c). These results strongly suggest that our theory is universal for models with long-range hopping.

For multi-chain/band cases, the bulk-bulk correspondence is also consistent with the prediction of our theory. If one set
\begin{equation}
U=U_1=diag[\beta,\beta^2,\beta^3,\cdots,\beta^{N},\beta,\beta^2,\cdots,\beta^{N}]
\end{equation}
the minimum sitting at $\beta=1$ for the considered model shown in Figs. \ref{Re1}(e) and (f), where the eigenvalues under $PBC$ and $OBC$ conditions roughly overlap. However, the NHSEs could not be correctly described since there is an even smaller value of $\mathcal{F}(\beta)$ for samples with finite-size by considering the following transformation:
\begin{equation}
U=U_2=diag[\beta_1,\beta_1^2,\beta_1^3,\cdots,\beta_1^{N},\beta_2,\beta_2^2,\cdots,\beta_2^{N}].
\end{equation}
Thus, the problem now falls into the multi-parameter optimization problems. As shown in Fig.\ref{Re2} (c), the global minimum satisfies $\beta_1=1/\beta_2$. It indicates the two chains have opposite directions of NHSE but with the same strength. Furthermore, $\beta_1$ approaches one with the increasing of $t_y$, indicating that $t_y$ suppress the strength of NHSE.
These results are consistent with the plot of $|\psi|^2$ shown in Fig. \ref{Re2}(b). One should also notice that the NHSEs under finite size is significantly suppressed by increasing $t_y$. Such a tendency is consistent with the variation of $\beta_1=1/\beta_2$ in Fig. \ref{Re2}(c) as well.

 We also notice that the winding number approach fails to capture the critical NHSEs \cite{NHSEs13}.
As shown in Fig. \ref{Re2}(e)
the standard winding number approach gives rise to a trivial winding number for sample size $N=60$, which implies the absence of NHSEs for samples with finite size. However, these results are contradict with the fact that the NHSEs are still available for $N=60$ [see Fig. \ref{Re2}(b)], where the eigenvalues for $H_{OBC}$ do not consist with those for $H_{PBC}$ [see Fig. \ref{Re2}(d)].

\subsection{B. Unique feature for the double-chain cases: coupling-reversed NHSE's directions}

\begin{figure}[h]
   \centering
    \includegraphics[width=0.75\textwidth]{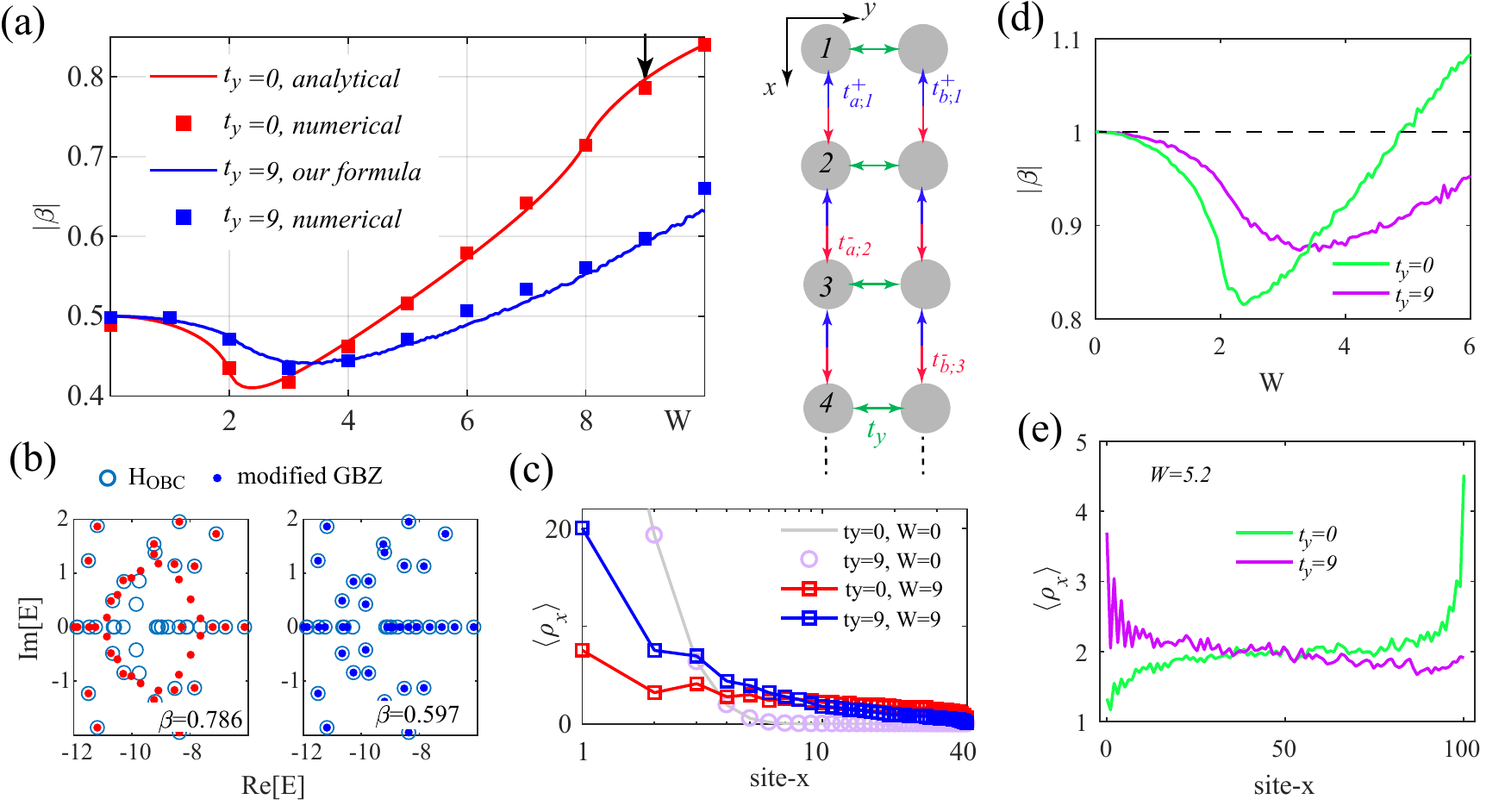}
    \caption{(Color online). The right panel in (a) gives the model. The left panel shows the plot of $|\beta|$ versus disorder strength $W$ for different $t_y$. The numerical results are obtained by using the proposed numerical approach. The solid lines are obtained by using the predicted analytical formulas. (b) The plots of eigenvalues for different transformation coefficient obtained by using the one-chain ($\widetilde{\beta}_{a,b}$) and double-chain formulas ($\widetilde{\beta}_d$). The disorder strength is $W=9$. Other parameters are $t^+_{a,b;i}=4+w^+_{a,b;i}$ and $t^-_{a,b;i}=1+w^-_{a,b;i}$.
     (c) The plot of $\rho_x=\sum_{i,y}|\psi_i(x,y)|^2$ for clean and disordered cases. (d) and (e) are the same with (a) and (c), except $t^-_{a,b;i}=1+w_{a,b;i}$ and $t^+_{a,b;i}=1$ with $w_{a,b;i}\in[-\frac{W}{2},\frac{W}{2}]$. The disorder strength $W$ has been marked in the figure. When disorder is considered, the predicted coefficient $\widetilde{\beta}_d$ successfully eliminates all the NHSEs, as shown in (b).
      }
   \label{Re2_2}
\end{figure}

We start from the model shown in Fig. \ref{Re2_2}(a). The Hamiltonian reads
\begin{equation}
\mathcal{H}=\sum_i-t^+_{a,b;i}c^\dagger_{a,b;i}c_{a,b;i+1}-
t^-_{a,b;i}c^\dagger_{a,b;i+1}c_{a,b;i}-t_yc^\dagger_{a;i}c_{b;i}-t_yc^\dagger_{b;i}c_{a;i}
\end{equation}
 where $t^\pm_{a,b;i}=t^{\pm}+w^{\pm}_{a,b;i}$ and $w^{\pm}_{a,b;i}\in[-\frac{W}{2},\frac{W}{2}]$. $a$ and $b$ stand for different chains.

On the restriction of $|t_y|> |t_{a,b;i}^{\pm}|$, we pay our attention to model shown in Fig. \ref{Re2_2}(a). $|t_y|> |t_{a,b;i}^{\pm}|$ is satisfied by setting $t_y=9$, $t^+=4$, $t^-=1$ and $W=9$. We have obtained an analytical formula in Sec. \B{VII. F} when $|t_y|> |t_{a,b;i}^{\pm}|$, where
\begin{equation}
\mathcal{F}(\beta)\approx  |\chi\prod_{i=1}^{N-1} [t_y^N\beta^{-N}(t_{a;i}^- + t_{b;i}^-)(t_{a;L}+t_{b;L})]+
\chi\prod_{i=1}^{N-1}[t_y^N\beta^{N}(t_{a;i}^+ + t_{b;i}^+)(t_{a;R}+t_{b;R})]|.
\end{equation}
$\chi$ is a constant. Here, we utilized $\beta_1=\beta_2=\beta$ since the exchanging of $a$ and $b$ chains do not influence the NHSE features. These terms are the largest leading term balancing the influence of $\beta$ and $t_y$.
Then, the required minimum gives [see Sec. \B{VII. F} for more details]:
\begin{equation}
\widetilde{\beta}_d^{2N}\approx \prod_{i=1}^{N} [\frac{t_y^N(t_{a;i}^- + t_{b;i}^-)}{t_y^N(t_{a;i}^+ + t_{b;i}^+)}]=\prod_{i=1}^{N} [\frac{2t^- + w_{a;i}^- + w_{b;i}^-}{2t^+ + w_{a;i}^+ + w_{b;i}^+}]
\label{EQbetad}
\end{equation}
The numerical results in Fig. \ref{Re2_2}(a) ensure the correctness of the considered formula, which is consistent with Eq. (\ref{EQbetad}). The correctness of our results is also identified by the consistency of the eigenvalues shown in Fig. \ref{Re2_2}(b).

One should notice that $(w_{a;i}^\pm + w_{b;i}^\pm)$ does not simply satisfy the uniform distribution $[-W,W]$ even $w_{a,b;i}^\pm\in[-\frac{W}{2},\frac{W}{2}]$.
Thus, some specific features can be identified:
(1). For clean samples with $t^{\pm}_{a;i}=t^\pm_{b;i}$, the $\widetilde{\beta}_d$ is the same with the Hatano-Nelson model, where $\widetilde{\beta}_{a,b}\approx[\prod_i \frac{t_{a,b;i}^-}{t_{a,b;i}^+ }]^{1/(2N)}$ and $\widetilde{\beta}_a=\widetilde{\beta}_b$, i.e., $\widetilde{\beta}_a=\widetilde{\beta}_b=\widetilde{\beta}_d$;
 (2). For disordered cases, $\widetilde{\beta}_a=\widetilde{\beta}_b$ still holds. However, the $\widetilde{\beta}_d$ is significantly distinct from $\widetilde{\beta}_{a,b}$ [see Figs. \ref{Re2_2}(a) and (b)].
Taking $W=9$ as an example, the NHSE show higher strength (higher stability against disorder) when $t_y=9$ since $\widetilde{\beta}_d<\widetilde{\beta}_{a,b}<1$ [see Fig. \ref{Re2_2}(c)].

More importantly, except for the quantitative differences, the coupled and decoupled cases also show qualitatively differences for disordered samples.
 By coupling two Hatano-Nelson chains each with NHSE in the right direction, the NHSE for the whole system is in the left direction instead, as shown in Figs. \ref{Re2_2}(d)-(e).
 For the decoupled case, both chain's NHSEs are in the right directions. At first glance, the NHSE's direction should be in the right direction when $t_y\neq0$, since one only simply couples the two chains with NHSEs both in the right directions.
However, based on our theory, the resulted NHSE's direction is in the left instead, as shown in Fig. \ref{Re2_2}(e).
 The coupling-induced reversing of NHSE's direction for the disordered samples can only be identified by our theory, which is unique for disordered non-Hermitian systems.

\subsection{C. Applicability to 2-D samples}

\begin{figure*}[h]
\includegraphics [width=14 cm]{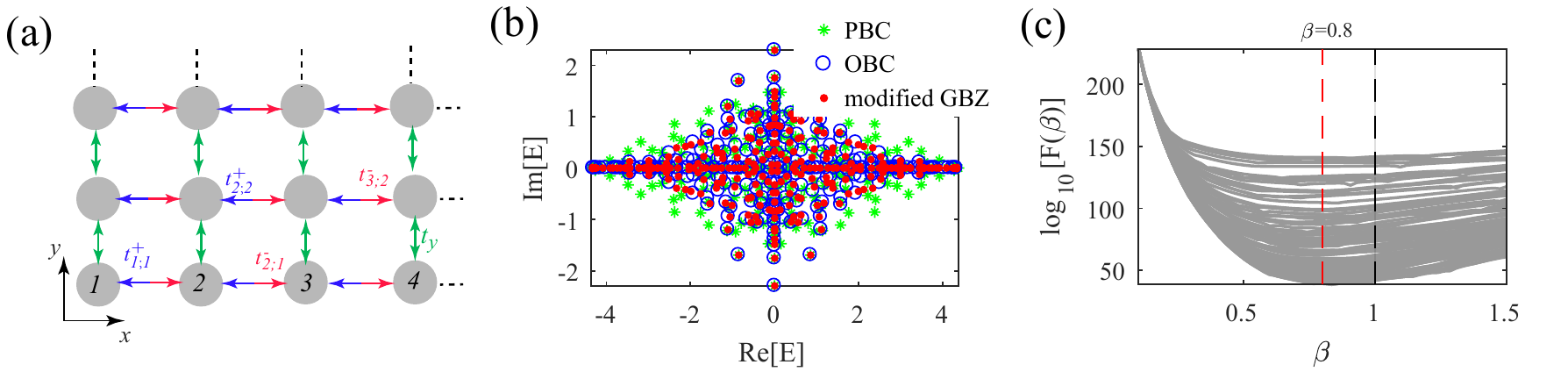}
\caption{(a) Scheme diagram of the two dimensional model. We set $N_x=16$ and $N_y=16$.
 The hopping is set as $t^\pm_{x,y}=t^{\pm}+w^\pm_{x,y}$ with $w^\pm_{x,y}\in[-W/2,W/2]$.
 For clean samples, one has $t^+=1.4$ and $t^-=0.6$ with $\beta_{\mathcal{GBZ}}\approx0.655$. We set $W=4$ and $t_y=1$.
(b) PBC, OBC and modified GBZ correspond to the eigenvalues of $\mathcal{H_{PBC}}$, $\mathcal{H_{OBC}}$ and $\mathcal{\widetilde{H}_{PBC}}(\beta)$. For $\mathcal{\widetilde{H}_{PBC}}(\beta)$, the $\beta=0.8$ is marked in (c).
(c) $\mathcal{F}(\beta)=|\det[E-\mathcal{\widetilde{H}_{PBC}}(\beta)]-\det[E-\mathcal{H_{OBC}}]|$ for different eigenvalues $E$ in (b). The slight deviations between modified GBZ and OBC in (b) can be further reduced by considering the energy dependent $\beta$. }
\label{FIGC1}
\end{figure*}

A two-dimensional system under disorder is also considered as shown in Fig. \ref{FIGC1}. The correctness of our approach for the BBC can be identified by comparing the eigenvalues for OBC and modified GBZ in Fig. \ref{FIGC1}(b). The $\beta$ for the modified GBZ is determined by the results in Fig. \ref{FIGC1}(c).

\subsection{D. Applicability to samples with on-site disorders}
The application of our approach for samples with on-site disorders is available in our study Ref. \cite{SROSMGBZ}

\subsection{E. Breaking the limitation of GBZ theory: off-diagonal similarity transformation}

The last example is related to an specific toy model shown in Fig. \ref{Re4}(a).
 Due to the requirement of translational symmetry for GBZ theory, the similarity transformation matrix $U$ to eliminate/generate the NHSEs is restricted to diagonal matrix \cite{SGBZ1}. Nevertheless, such a requirement is unnecessary for samples without translational symmetry in the frame of our theory [see Fig. 1(b) in the main text]. For instance, one can use
 \begin{equation}
 V=\left[
 \begin{array}{cccccc}
 1 & \beta &0&0&0&\cdots\\
 0&1&\beta&0&0&\cdots\\
 0&0&1&\beta&0&\cdots\\
 \vdots&\vdots&\ddots&\ddots&\ddots&\cdots
 \end{array}
 \right].
 \label{EQVt}
 \end{equation}
The transformation matrix $V$ has distinct features compared with those in GBZ theory.

We pay our attention to the Hamiltonian $\mathcal{H}_{\kappa}=\sum_{i}t^+c^\dagger_i c_{i+1}+t^-c^\dagger_{i+1} c_{i}$ with  $t^\pm=1$ with extended eigenstates $|\psi_i\rangle$.
 Although $V|\psi_i\rangle$ does not show the NHSE, $V^{-1}|\psi_i\rangle$ possesses the NHSE. It is distinct from those in GBZ theory. The transformation matrix $U=diag[\beta,\beta^2,\beta^3,\cdots]$ guarantees the existence of NHSEs for both $U|\psi_i\rangle$ and $U^{-1}|\psi_i\rangle$ when $\beta\neq1$. Specifically, the matrix $\mathcal{H}_s=V\mathcal{H}_{\kappa}V^{-1}$ and its transpose $[\mathcal{H}_s]^T$ do not have the NHSEs at the same time even the asymmetry hopping still exists for both of them. It can be considered as the first unique features for $V$ matrix predicted by our theory.

 \begin{figure}[h]
   \centering
    \includegraphics[width=0.75\textwidth]{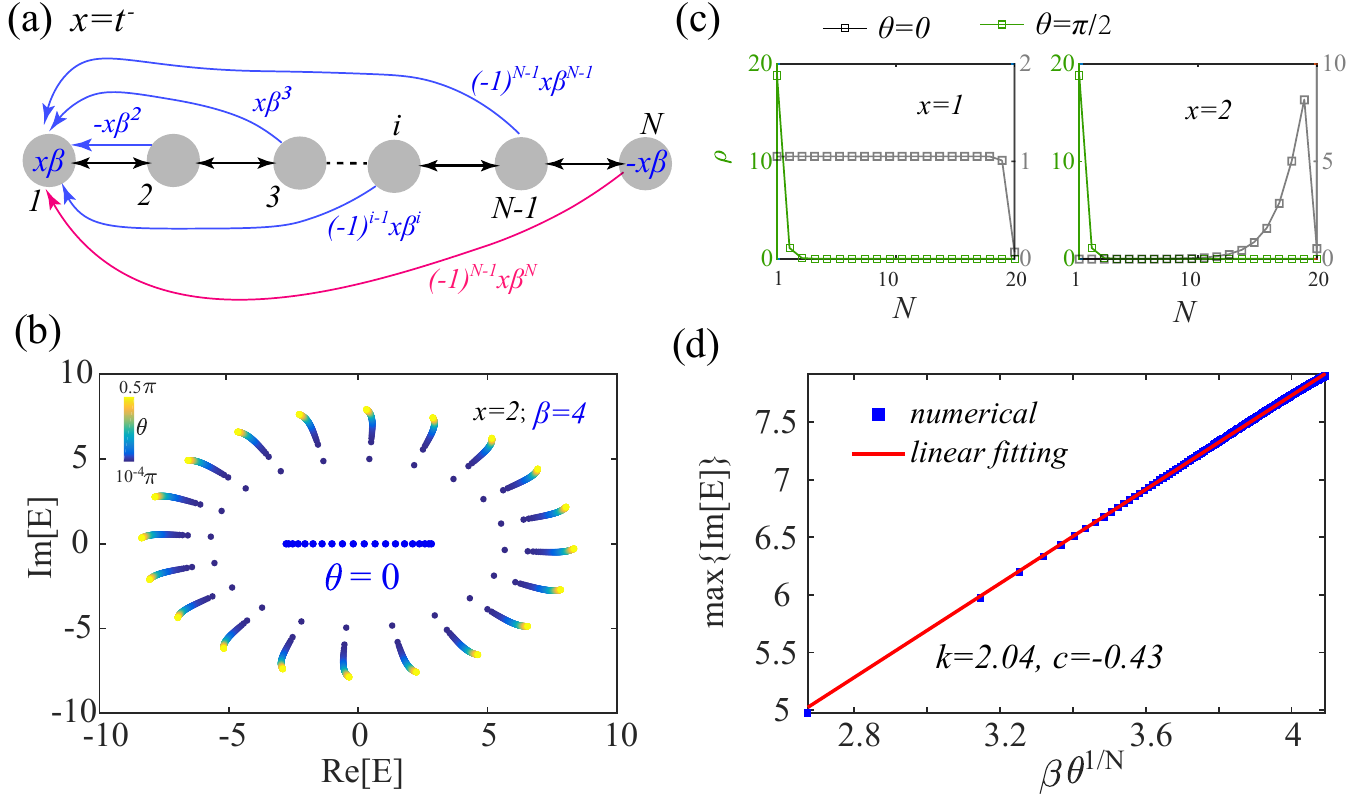}
    \caption{(Color online).
     The Hamiltonian in (a) reads $V\mathcal{H}_{\kappa}V^{-1}$, where $\mathcal{H}_{\kappa}=\sum_{i}t^+c^\dagger_i c_{i+1}+t^-c^\dagger_{i+1} c_{i}$. $V$ is given in Eq. (\ref{EQVt}). The parameters are $t_{i}^-=x$, $t^+_i=1$ and $\beta=4$.
     (b) The plot of $\rho$ for different $\theta$ for model in (a). (c) The plot of $Re[E]$ versus $Im[E]$ for different $\theta$. (d) The linear fitting of $max\{Im[E]\}$ versus $\theta$ with $max\{Im[E]\}=k\beta\theta^{1/N}+c$. The data in (b) is adopted.   }
   \label{Re4}
\end{figure}

For model $\mathcal{H}_s=V\mathcal{H}_{\kappa}V^{-1}$ [see Fig. \ref{Re4} (a)], we also found that a pure flux $e^{j\theta}$ applied on hopping between the $1_{th}$-site and the $N_{th}$-site can be used to modulate the NHSEs, as shown in Fig. \ref{Re4}(c).
The distribution of eigenvalues is also changed from a critical line to a closed-loop. Furthermore, the variation of the eigenvalues captures the generalized scaling features $max\{Im[E]\}\propto x\beta\theta^{1/N}$ with $x=t^-$, as shown in Figs. \ref{Re4}(b) and (d).
 These $\theta$ dependent features do not exist in the Hatano-Nelson model, and neither can be understood by any other non-Hermitian theory. Nevertheless, it can be simply understood by our proposed scheme as follows.

 Firstly, $\mathcal{H}_s$ should has the same determinant with $\mathcal{H}_{\kappa}$ and is $\beta$ independent, with
 $\det[E-\mathcal{H}_s]=\det[E-\mathcal{H}_{\kappa}]$.
 By considering a phase $e^{j\theta}$ between $1_{th}$ and $N_{th}$ sites. The determinant is obtained:
  \begin{equation}
 \det[E-\mathcal{H}_s(\theta)]=\det[E-\mathcal{H}_{\kappa}]+(-1)^{N}x^N\beta^Ne^{j\theta} +(-1)^{N+1}x^N\beta^N.
 \end{equation}
A difference [$(-x\beta)^N(1-e^{j\theta})$] between $\mathcal{H}_s(\theta)$ and $\mathcal{H}_{\kappa}$ exists, which is zero for $\theta=0$.
Based on our theory, such a model can be mapped to $\mathcal{H}_{\kappa}$ with periodic boundary condition $t_L=0$ and $t_R=(-1)^{N}x\beta^N(1-e^{i\theta})$ \cite{CS7a}, since they have the same determinant. It indicates that one can verify the NHSEs and the eigenvalues by simply modulating $\theta$ in general.
 In short, the unique features of such a specific model can also be predicted and clarified by our theory.

\section{ Details of the deviations and proofs}

\subsection{A. Determinant of the Hatano-Nelson model under open and periodic boundary conditions}

It is known that the eigenvalues calculated under open and periodic boundary conditions are different when samples are in small sizes. To achieve reliable $BBC$, large sample size with $n\rightarrow\infty$ should be considered.
In this section, we present the analytical results for samples in which $n\rightarrow \infty$ while the translational symmetry is absent. The main results are independent of $n$ when $n\rightarrow\infty$. For simplicity, we set $n=4k$ as an even number throughout the deviations only.

\subsubsection{1. Determinant under open boundary condition: $\det[E-\mathcal{H}_{\mathcal{OBC};~n\times n}]$}
We focus on the disordered Hatano-Nelson model \cite{SHN} with Hamiltonian:
\begin{align}
\begin{split}
\mathcal{H} =\sum_{i=1}^{n} -t^+_ic_{i}^\dagger c_{i+1}-t^-_{i}c_{i+1}^\dagger c_{i}.
\label{EQSmd1}
\end{split}
\end{align}
The determinant $\det[E-\mathcal{H}_{\mathcal{OBC};~n\times n}]$ under OBC has the form of:
\begin{equation}
\det [E-\mathcal{H}_{\mathcal{OBC};~n\times n}] =
\left|
\begin{array}{cccccccccccccccc}
E   & t_1^+ & 0   & 0   & 0    & 0   & 0 & 0        & \cdots   & 0  \\
t_1^- & E   & t_2^+ & 0   & 0    & 0   & 0 & 0        & \cdots   & 0     \\
0   & t_2^- & E   & t_3^+ & 0    & 0   & 0 & 0        & \cdots   & 0   \\
0   & 0   & t_3^- & E   & t_4^+  & 0   & 0 & 0        & \cdots   & 0   \\
0   & 0   & 0   & t_4^- & E    & t_5^+ & 0 & 0        & \cdots   & 0  \\
0   & 0   & 0   & 0   & t_{5}^- & E   & t_{6}^+  & 0        & \cdots   & 0     \\
\vdots   & \vdots   & \vdots   & \vdots   & \vdots    & \ddots   & \ddots & \ddots        & \ddots   & 0   \\
0   & 0   & 0   & 0   & 0    & 0   & 0 & E        & t_{n-2}^+ & 0   \\
0   & 0   & 0   & 0   & 0    & 0   & 0 & t_{n-2}^- & E        & t_{n-1}^+   \\
0   & 0   & 0   & 0   & 0    & 0   & 0 & 0        & t_{n-1}^-   & E   \\
\end{array}
\right|_{n\times n}.
\label{EQhn}
\end{equation}
Generally, the recurrence relation
\begin{align}
\begin{split}
\det [E-\mathcal{H}_{\mathcal{OBC};~n\times n}]=\det [E-\mathcal{H}_{\mathcal{OBC};~n-1\times n-1}]E-t^+_1\Delta_x,
\end{split}
\end{align}
can be adopted to evaluate the determinant, where the $\Delta_x$ reads
\begin{equation}
\Delta_x =
\left|
\begin{array}{cccccccccccccccc}
t_1^-  & t_2^+ & 0   & 0    & 0   & 0 & 0        & \cdots   & 0     \\
0    & E   & t_3^+ & 0    & 0   & 0 & 0        & \cdots   & 0   \\
0    & t_3^- & E   & t_4^+  & 0   & 0 & 0        & \cdots   & 0   \\
0    & 0   & t_4^- & E    & t_5^+ & 0 & 0        & \cdots   & 0  \\
0    & 0   & 0   & t_{5}^- & E   & t_{6}^+  & 0        & \cdots   & 0     \\
\vdots     & \vdots   & \vdots   & \vdots    & \ddots   & \ddots & \ddots        & \ddots   & 0   \\
0      & 0   & 0   & 0    & 0   & 0 & E        & t_{n-2}^+ & 0   \\
0      & 0   & 0   & 0    & 0   & 0 & t_{n-2}^- & E        & t_{n-1}^+   \\
0      & 0   & 0   & 0    & 0   & 0 & 0        & t_{n-1}^-   & E   \\
\end{array}
\right|_{n-1\times n-1}.
\end{equation}
Hence, $\Delta_x=t_1^- \det [E-\mathcal{H}_{\mathcal{OBC};~n-2\times n-2}]$ and then one has
\begin{align}
\begin{split}
\det [E-\mathcal{H}_{\mathcal{OBC};~n\times n}]=\det [E-\mathcal{H}_{\mathcal{OBC};~n-1\times n-1}]E-t_1^+t_1^- \det [E-\mathcal{H}_{\mathcal{OBC};~n-2\times n-2}].
\end{split}
\end{align}
Similarly we have:
\begin{align}
\begin{split}
&\det [E-\mathcal{H}_{\mathcal{OBC};~n-1\times n-1}]=\det [E-\mathcal{H}_{\mathcal{OBC};~n-2\times n-2}]E-t_2^+t_2^- \det [E-\mathcal{H}_{\mathcal{OBC};~n-3\times n-3}],\\
&\det [E-\mathcal{H}_{\mathcal{OBC};~n-2\times n-2}]=\det [E-\mathcal{H}_{\mathcal{OBC};~n-3\times n-3}]E-t_3^+t_3^- \det [E-\mathcal{H}_{\mathcal{OBC};~n-4\times n-4}].
\end{split}
\end{align}
In such a way, the determinant can be simply marked as:
\begin{align}
\begin{split}
\det [E-\mathcal{H}_{\mathcal{OBC};~n\times n}]=
\sum\limits_{m=0}^{n/2}(-1)^{m}[\sum_i\prod\limits_{i\in[1,n-1]}^{m~pair}t_i^+t_{i}^-]
E^{n-2m}= f(t_i^+t_{i}^-,E).
\label{EQS23}
\end{split}
\end{align}
$m~pair$ means that one has to select $m$ pair of $t^+_it^-_{i}$ with $i\equiv[i_1,i_2,\cdots,i_x\cdots, i_y,\cdots,i_m]\in[1,n-1]$. For simplicity, we mark it as $m$ here after. The summation $\sum_i$ suggests that one has to include all the possible permutation of $i\equiv[i_1,i_2,\cdots,i_x,\cdots, i_y,\cdots,i_m]$ for fixed $m$, where the restriction $|i_x-i_y|\geq2$ for $\forall$ $i_x,i_y\in[1,n-1]$ should be considered at the same time.
In specific, we give the detailed formula as follow. Eq. (\ref{EQS23}) can be rewritten as:
\begin{align}
\begin{split}
\det [E-\mathcal{H}_{\mathcal{OBC};~n\times n}]
&=E^n
 -\sum_{i_1=1}^{n-1}\mathcal{T}_{(n-i_1)}E^{n-2}
 +\sum_{i_1=1}^{n-3}\mathcal{T}_{(n-i_1)}\sum_{i_2=i_1+2}^{n-1}\mathcal{T}_{(n-i_2)}E^{n-4}\\
&-\sum_{i_1=1}^{n-5}\mathcal{T}_{(n-i_1)}\sum_{i_2=i_1+2}^{n-3}\mathcal{T}_{(n-i_2)}\sum_{i_3=i_2+2}^{n-1}\mathcal{T}_{(n-i_3)}E^{n-6}+\cdots\\
&+(-1)^{n/2}\sum_{i_1=1}^{1}\mathcal{T}_{(n-i_1)}\sum_{i_2=i_1+2}^{3}\mathcal{T}_{(n-i_2)}\sum_{i_3=i_2+2}^{5}\mathcal{T}_{(n-i_3)}\cdots
 \sum_{i_{\frac{n}{2}}=i_{\frac{n-2}{2}-1}+2}^{n-1}\mathcal{T}_{(n-i_{\frac{n-2}{2}})} E^0
\end{split}
\end{align}
with the mark $\mathcal{T}_i=t^-_{i}t^+_{i}$.

Actually, the detailed expression of $\det [E-\mathcal{H}_{\mathcal{OBC};~n\times n}]$ is unimportant. Instead, the important thing is that such a determinant only contains $\mathcal{T}_i=t^+_it^-_{i}$ and $E^2$ terms for $OBC$, while the isolated $t_i^+$ or $t_i^-$ term is absent. Such a fact is very important to achieve a specific transformation preserving $\det [E-\mathcal{H}_{\mathcal{OBC};~n\times n}]$. One only needs to ensure that $t_i^+t_{i}^-$ remains unchanged for $ \forall ~i\in[1,n-1]$. For the sake of convenience, in the following derivation, the corresponding determinant is expressed by the simplified formula as Eq. (\ref{EQS23}).

\subsubsection{2. Determinant under periodic boundary condition: $\det[E-\mathcal{H}_{\mathcal{PBC};~n\times n}]$}

The matrix form of Hamiltonian in the main text under $PBC$ is given:
\begin{equation}
E-\mathcal{H}_{\mathcal{PBC};~n\times n} =
\left(
\begin{array}{cccccccccccccccc}
E   & t_1^+ & 0   & 0   & 0    & 0   & 0 & 0        & \cdots   & t_L  \\
t_1^- & E   & t_2^+ & 0   & 0    & 0   & 0 & 0        & \cdots   & 0     \\
0   & t_2^- & E   & t_3^+ & 0    & 0   & 0 & 0        & \cdots   & 0   \\
0   & 0   & t_3^- & E   & t_4^+  & 0   & 0 & 0        & \cdots   & 0   \\
0   & 0   & 0   & t_4^- & E    & t_5^+ & 0 & 0        & \cdots   & 0  \\
0   & 0   & 0   & 0   & t_{5}^- & E   & t_{6}^+  & 0        & \cdots   & 0     \\
\vdots   & \vdots   & \vdots   & \vdots   & \vdots    & \ddots   & \ddots & \ddots        & \ddots   & 0   \\
0   & 0   & 0   & 0   & 0    & 0   & 0 & E        & t_{n-2}^+ & 0   \\
0   & 0   & 0   & 0   & 0    & 0   & 0 & t_{n-2}^- & E        & t_{n-1}^+   \\
t_R   & 0   & 0   & 0   & 0    & 0   & 0 & 0        & t_{n-1}^-   & E   \\
\end{array}
\right)_{n\times n}.
\end{equation}
with $t^+_{n}=t_R\equiv (t+\gamma+w^+_n)$ and $t^-_{n}=t_L\equiv (t-\gamma+w^-_{n})$.
The determinant can be rewritten as:
\begin{align}
\begin{split}
 \det[E-\mathcal{H}_{\mathcal{PBC};~n\times n}]=\det[E-\mathcal{H}_{\mathcal{OBC};~n\times n}]+f(t_L,t_R) ,
\label{EQS14}
\end{split}
\end{align}
with \cite{SLNAB}
\begin{align}
\begin{split}
f(t_L,t_R;E)=&
 (-1)^{\tau[n;2,\cdots,n-1;1]}t_Lt_R\det[E-\mathcal{H}_{\mathcal{OBC};~n-2\times n-2}]+\\
&(-1)^{\tau[n;1,\cdots,n-1]}t_Lt_1^-t_2^-\cdots t_{m}^-\cdots t_{n-1}^-+\\
&(-1)^{\tau[2,\cdots,n;1]}t_Rt_1^+t_2^+\cdots t_{m}^+\cdots t_{n-1}^+.
\label{EQS15}
\end{split}
\end{align}
The permutation of order $n$ is in the form of:
\begin{align}
\begin{split}
(-1)^{\tau[n;(2,\cdots,n-1);1]}&=(-1)^{[(n-1)+(n-2)]}=-1;\\
(-1)^{\tau[n;1,\cdots,n-1]}&=(-1)^{n-1}=-1;\\
(-1)^{\tau[2,\cdots,n;1]}&=(-1)^{n-1}=-1;
\end{split}
\end{align}
with $n=4k$. Thus,
\begin{align}
\begin{split}
f(t_L,t_R;E)=&
 -t_Lt_R\det[E-\mathcal{H}_{\mathcal{OBC};~n-2\times n-2}]+\\
&-t_Lt_1^-t_2^-\cdots t_{m}^-\cdots t_{n-1}^-\\
&-t_Rt_1^+t_2^+\cdots t_{m}^+\cdots t_{n-1}^+.
\end{split}
\end{align}
Based on Eq. (\ref{EQS23}), one has
\begin{align}
\begin{split}
\det [E-\mathcal{H}_{\mathcal{OBC};~n-2\times n-2}]=
\sum\limits_{m=0}^{n/2-1}(-1)^{m}[\sum_i\prod\limits_{i\in[2,n-2]}^mt_i^+t_{i}^-]E^{(n-2)-2m}.
\end{split}
\end{align}
Then, Eq. (\ref{EQS15}) can be written as
\begin{align}
\begin{split}
 f(t_L,t_R;E)=
&-t_Lt_R\{\sum\limits_{m=0}^{n/2-1}(-1)^{m}[\sum_i\prod\limits_{i\in[2,n-2]}^mt^+_it^-_{i}]E^{(n-2)-2m}\}\\
&-t_Lt_1^-t_2^-\cdots t_{m}^-\cdots t_{n-1}^-\\
&-t_Rt_1^+t_2^+\cdots t_{m}^+\cdots t_{n-1}^+.
\end{split}
\end{align}
Based on these results, Eq. (\ref{EQS14}) is given as follows:
\begin{align}
\begin{split}
\det[E-\mathcal{H}_{\mathcal{PBC};~n\times n}]
=&\det[E-\mathcal{H}_{\mathcal{OBC};~n\times n}]+f(t_L,t_R) \\
=&\{\sum\limits_{m=0}^{n/2}(-1)^{m}[\sum_i\prod\limits_{i\in[1,n-1]}^mt^+_it^-_{i}]E^{n-2m}\}
-t_Lt_R\{\sum\limits_{m=0}^{n/2-1}(-1)^{m}[\sum_i\prod\limits_{i\in[2,n-2]}^mt^+_it^-_{i}]E^{(n-2)-2m}\}\\
&-[t_Lt_1^-t_2^-\cdots t_{m}^-\cdots t_{n-1}^-]-[t_Rt_1^+t_2^+\cdots t_{m}^+\cdots t_{n-1}^+].
\label{EQS26}
\end{split}
\end{align}

\subsection{B. Deviations of the requirements of BBC for the Hatano-Nelson model }\label{section5}

In this section, we demonstrate how to rebuild the BBC in disordered non-Hermitian systems with nearest-neighbor hopping.
For a disordered $\mathcal{H}$, the determinant under $PBC$ reads:
\begin{align}
\begin{split}
\det[E-\mathcal{H}_{\mathcal{PBC};~n\times n}]
=&\det[E-\mathcal{H}_{\mathcal{OBC};~n\times n}] -t_Lt_R\det[E-\mathcal{H}_{\mathcal{OBC};~n-2\times n-2}] \\
&-[t_Lt_1^-t_2^-\cdots t_{m}^-\cdots t_{n-1}^-]-[t_Rt_1^+t_2^+\cdots t_{m}^+\cdots t_{n-1}^+].
\label{EQSx}
\end{split}
\end{align}
For a specific eigenvalue $E_{\mathcal{OBC}}$ belonging to the open boundary spectrum, one should have
 \begin{align}
\begin{split}
\det[E_{\mathcal{OBC}}-\mathcal{H}_{\mathcal{OBC};~n-2\times n-2}]=0.
\end{split}
\end{align}
If $n\rightarrow \infty$, it is reasonable to demonstrate that
 \begin{align}
\begin{split}
\det[E_{\mathcal{OBC}}-\mathcal{H}_{\mathcal{OBC};~n\times n}]\approx0.
\label{EQSKK}
\end{split}
\end{align}
The proof is given in the next subsection. Thus, for systems with non-Hermitian skin effect (NHSE), one has:
\begin{align}
\begin{split}
\det[E_{\mathcal{OBC}}-\mathcal{H}_{\mathcal{PBC};~n\times n}]
=&\det[E_{\mathcal{OBC}}-\mathcal{H}_{\mathcal{OBC};~n\times n}] -t_Lt_R\det[E_{\mathcal{OBC}}-\mathcal{H}_{\mathcal{OBC};~n-2\times n-2}] \\
&-[t_Lt_1^-t_2^-\cdots t_{m}^-\cdots t_{n-1}^-]-[t_Rt_1^+t_2^+\cdots t_{m}^+\cdots t_{n-1}^+]\\
\approx & 0-0-[t_Lt_1^-t_2^-\cdots t_{m}^-\cdots t_{n-1}^-]-[t_Rt_1^+t_2^+\cdots t_{m}^+\cdots t_{n-1}^+].
\end{split}
\end{align}
Suppose such an eigenvalue also belongs to the spectrum under the $PBC$, one should have
 \begin{align}
\begin{split}
\det[E_{\mathcal{OBC}}-\mathcal{H}_{\mathcal{PBC};~n\times n}]\approx-[t_Lt_1^-t_2^-\cdots t_{m}^-\cdots t_{n-1}^-]-[t_Rt_1^+t_2^+\cdots t_{m}^+\cdots t_{n-1}^+]\approx0,
\end{split}
\end{align}
However, it is not the case for $\mathcal{H}$.
Nevertheless, by introducing a transformation $H\rightarrow\widetilde{\mathcal{H}}$, $E_{\mathcal{OBC}}$ could be approximately the eigenvalues of $\widetilde{\mathcal{H}}$ under $PBC$.
Now one can see what is the requirement to obtain $\det[E_{\mathcal{OBC}}-\mathcal{\widetilde{H}}_{\mathcal{\mathcal{PBC}};~n\times n}]\rightarrow 0$.
 Generally, there are two key issues:

(i) One has to ensure that the transformation preserves the following equation: $\det[E_{\mathcal{OBC}}-\mathcal{\widetilde{H}}_{\mathcal{\mathcal{OBC}};~n-2\times n-2}]=\det[E_{\mathcal{OBC}}-\mathcal{H}_{\mathcal{OBC};~n-2\times n-2}]=0$ and $\det[E_{\mathcal{OBC}}-\mathcal{\widetilde{H}}_{\mathcal{\mathcal{OBC}};~n\times n}]=\det[E_{\mathcal{OBC}}-\mathcal{H}_{\mathcal{OBC};~n\times n}]$.

(ii) One has to minimize $|[t_Lt_1^-t_2^-\cdots t_{m}^-\cdots t_{n-1}^-]+[t_Rt_1^+t_2^+\cdots t_{m}^+\cdots t_{n-1}^+]|$ to ensure the accuracy of $\det[E_{\mathcal{OBC}}-\widetilde{\mathcal{H}}_{\mathcal{PBC};~n\times n}]\rightarrow 0$.

Thus, $t^+_it^-_{i}$ should be fixed to preserve the first requirement, for instance $ t^+_i\rightarrow\widetilde{\beta} t^+_i$ and $ t^-_{i}\rightarrow\widetilde{\beta}^{-1} t^-_{i}$. As for the second key issue, one has to minimize $|f_x|$ with:
\begin{align}
\begin{split}
f_x=
\widetilde{\beta}^n[t_Rt^+_1t^+_2\cdots t_{m}^+\cdots t_{n-1}^+]
+\widetilde{\beta}^{-n} [t_Lt_1^-t_2^-\cdots t_{m}^-\cdots t_{n-1}^-]=
xT^++x^{-1}T^-,
\label{EQSfx}
\end{split}
\end{align}
and $x=\widetilde{\beta}^n$. With the variation of $x$, the minimum is available when $\frac{df_x}{dx}=0$. The validity of our results is confirmed numerically, see Sec. III for details. Then, one has $T^+-T^-x^{-2}=0$ with
\begin{align}
\begin{split}
x^2=\widetilde{\beta}^{2n}=\frac{t_Lt_1^-t_2^-\cdots t_{m}^-\cdots t_{n-1}^-}{t_Rt_1^+t_2^+\cdots t_{m}^+\cdots t_{n-1}^+}.
\label{EQSd}
\end{split}
\end{align}
If $t^+t^-$ is invariant when $\beta$ is in the form shown in Eq. (\ref{EQSd}), then $\det[E_{\mathcal{OBC}}-\mathcal{\widetilde{H}}_{\mathcal{\mathcal{PBC}};~n\times n}]\rightarrow0$ holds to the maximum extent. Thus, all of $\mathcal{\widetilde{H}}_{\mathcal{\mathcal{PBC}}}$, $\mathcal{\widetilde{H}}_{\mathcal{\mathcal{OBC}}}$ and $\mathcal{H}_{\mathcal{OBC}}$ tend to have the same eigenvalue $E_{\mathcal{OBC}}$.

In short, one has
\begin{align}
\begin{split}
f_x=&\widetilde{\beta}^n[t_Rt_1^+t_2^+\cdots t_{m}^+\cdots t_{n-1}^+]+\widetilde{\beta}^{-n}[t_Lt_1^-t_2^-\cdots t_{m}^-\cdots t_{n-1}^-]\\
=&2\sqrt{[t_Lt_1^-t_2^-\cdots t_{m}^-\cdots t_{n-1}^-][t_Rt_1^+t_2^+\cdots t_{m}^+\cdots t_{n-1}^+]}.
\end{split}
\end{align}
Such a result is also consistent with the transformation $\beta_i\equiv\sqrt{t^-_{i}/t^+_i}$ with $\widetilde{t}^+_i=\beta_it^+_i$ and $\widetilde{t}^-_{i}=\beta^{-1}_it^-_{i}$, where
\begin{align}
\begin{split}
f_x=&\prod_i\beta_i[t_Rt_1^+t_2^+\cdots t_{m}^+\cdots t_{n-1}^+]+\prod_i\beta_i^{-1}[t_Lt_1^-t_2^-\cdots t_{m}^-\cdots t_{n-1}^-]\\
=&2\sqrt{[t_Lt_1^-t_2^-\cdots t_{m}^-\cdots t_{n-1}^-][t_Rt_1^+t_2^+\cdots t_{m}^+\cdots t_{n-1}^+]}.
\end{split}
\end{align}

\subsection{C. Plots of Eq. (\ref{EQSfx}) }

 \begin{figure*}[h]
\includegraphics[width=16cm]{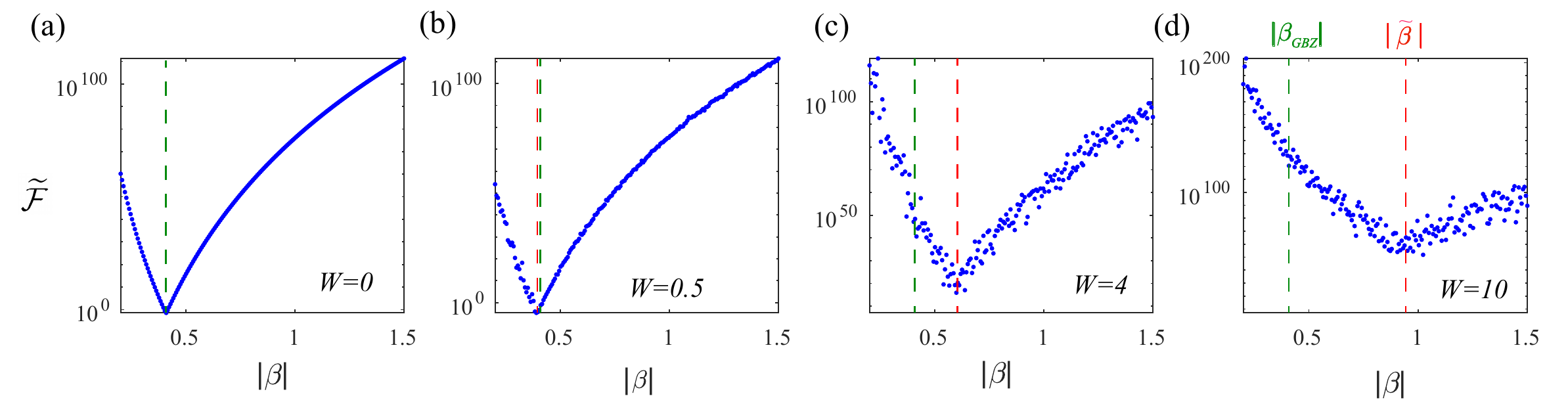}
\caption{(Color online).  The plots of $\widetilde{\mathcal{F}}\equiv|f(\beta)|=
|\beta^n[t_Rt^+_1t^+_2\cdots t_{m}^+\cdots t_{n-1}^+]
+\beta^{-n} [t_Lt_1^-t_2^-\cdots t_{m}^-\cdots t_{n-1}^-]|$. The hopping strength is $t^\pm_i=t\pm\gamma+w_i^{\pm}$ with $w_i^{\pm}\in[-\frac{W}{2},\frac{W}{2}]$. $W$ is the disorder strength. The green and red dashed line corresponds to the value of $|\beta_{\mathcal{GBZ}}|$ and $|\widetilde{\beta}|$, respectively. We set $n=200$. The disorder strength has been marked in the figures. The other parameters are $t=1$ and $\gamma=1.4$.
 }
\label{SF}
\end{figure*}

The plots of $\widetilde{\mathcal{F}}\equiv|f(\beta)|=
|\beta^n[t_Rt^+_1t^+_2\cdots t_{m}^+\cdots t_{n-1}^+]
+\beta^{-n} [t_Lt_1^-t_2^-\cdots t_{m}^-\cdots t_{n-1}^-]|$ in Eq. (\ref{EQSfx}) for different disorder strengths $W$ are given in Fig. \ref{SF}. One can see that the minimum of $\widetilde{\mathcal{F}}$ are consistent with our analytical results $\widetilde{\beta}$. For week disorder strength, one has $|\beta_{\mathcal{GBZ}}|\approx|\widetilde{\beta}|$, which is located at the minimum of $\widetilde{\mathcal{F}}$. However, when disorder is strong enough, $|\beta_{\mathcal{GBZ}}|\neq|\widetilde{\beta}|$.

\subsection{D. Equation (\ref{EQSKK}) is an appropriate approximation}

In this subsection, we prove that if $\det[E_{\mathcal{OBC}}-\mathcal{H}_{\mathcal{OBC};~n-2\times n-2}]=0$ and $n\rightarrow \infty$, then
 \begin{align}
\begin{split}
\det[E_{\mathcal{OBC}}-\mathcal{H}_{\mathcal{OBC};~n\times n}]\approx0.
\end{split}
\end{align}
is an appropriate approximation.

Since $\mathcal{H}_{\mathcal{OBC}}(t^+_i,t^-_{i})$ has the same eigenvalues with $\mathcal{\widetilde{H}}_{\mathcal{OBC}}(\beta_it^+_i,\beta_i^{-1}t^-_{i})$, we pay our attention to $\mathcal{\widetilde{H}}_{\mathcal{OBC}}(\beta_it^+_i,\beta_i^{-1}t^-_{i})$, in which $\beta_it^+_i=\beta_i^{-1}t^-_{i}=\sqrt{t^+_it^-_{i}}$ is a constant. $\mathcal{\widetilde{H}}_{\mathcal{OBC}}(\beta_it^+_i,\beta_i^{-1}t^-_{i})$ satisfies the eigenvalue equation:
\begin{align}
\begin{split}
 \mathcal{\widetilde{H}}_{\mathcal{OBC},n-2\times n-2}V=V E_\mathcal{OBC}.
\end{split}
\end{align}
Noticing that $\mathcal{\widetilde{H}}_{\mathcal{OBC},n-2\times n-2}$ has no asymmetric hopping, therefore the NHSE is absent based on our theory. The above equation can be rewritten as $V^{-1}\mathcal{\widetilde{H}}_{\mathcal{OBC},n-2\times n-2}V=E_\mathcal{OBC}$. Consider the following transformation:
 \begin{align}
\begin{split}
U^{-1}\mathcal{\widetilde{H}}_{\mathcal{OBC},n\times n}U&=
\left(\begin{array}{ccc}
1 & 0& 0\\
0 & V^{-1}& 0\\
0 & 0& 1
\end{array}\right)
\left(\begin{array}{ccc}
0 & \beta_1T^+_1& 0\\
\beta_1^{-1}T^-_1 & \mathcal{\widetilde{H}}_{\mathcal{OBC},n-2\times n-2}& \beta_2T_2^+\\
0 & \beta_2^{-1}T_2^-& 0
\end{array}\right)
\left(\begin{array}{ccc}
1 & 0& 0\\
0 & V& 0\\
0 & 0& 1
\end{array}\right)\\
&=
\left(\begin{array}{ccc}
0 & \beta_1T^+_1V& 0\\
\beta_1^{-1}V^{-1}T^-_1 & E_\mathcal{OBC}& \beta_2V^{-1}T_2^+\\
0 & \beta_2^{-1}T_2^-V& 0
\end{array}\right).
\end{split}
\end{align}
where $U$ satisfying $U^{-1}U=\mathbb{I}$ does not change the eigenvalues. Furthermore, one has  $T_{1}^{\pm}\equiv[t^\pm_{1},0,\cdots,0]_{n-2}$ and $T_{2}^{\pm}\equiv[0,0,\cdots,t^\pm_{2}]_{n-2}$, which lead to the expression of $(\beta_1T_{1}^+V)_k=\beta_1t_{1}^+V_{1,k}$ and $(\beta^{-1}_2T_{2}^-V)_k=\beta^{-1}_2t_{2}^-V_{n-2,k}$.

 Generally, the extended states are equally distributed in the lattice, which means that the eigenvector $V$ satisfies $V_{1,k}\propto \frac{1}{n-2}$ and $V_{n-2,k}\propto \frac{1}{n-2}$. For a localized state, instead, one has $V_{1,(n-2);k}\sim 0$ since the eigenvalues do not tend to concentrate at the first/last site for a sample with size $(n-2)$.
 Based on {\it Gershgorin circle theorem} \cite{SGCT} and $\beta_it^+_i\equiv constant$,
 the eigenvalues of $\mathcal{\widetilde{H}}_{\mathcal{OBC},n-2\times n-2}$ [$E_{k,n-2}\in E_\mathcal{OBC}$] are related to
 the eigenvalues of $\mathcal{\widetilde{H}}_{\mathcal{OBC},n\times n}$ [$E_{k,n}$] as
\begin{align}
\begin{split}
\lim\limits_{n\rightarrow\infty}[|E_{k,n}-E_{k,n-2}|]<\lim\limits_{n\rightarrow\infty}|\beta_1t_1^+V_{1,k}+\beta^{-1}_2t_{2}^-V_{n-2,k}|\propto \lim\limits_{n\rightarrow\infty}|\frac{\beta_1t_1^++\beta^{-1}_2t^-_2}{n-2}|\approx0.
\end{split}
\end{align}
It suggests that $E_{k,n}\approx E_{k,n-2}\in E_{\mathcal{OBC}}$ is also approximately the eigenvalue of $\mathcal{\widetilde{H}}_{\mathcal{OBC},n\times n}$.

Thus, if $\det[E_{\mathcal{OBC}}-\mathcal{\widetilde{H}}_{\mathcal{OBC};~n-2\times n-2}]=0$, one has $\det[E_{\mathcal{OBC}}-\mathcal{\widetilde{H}}_{\mathcal{OBC};~n\times n}]\approx0$.
Noticing that $\det[E-\mathcal{H}_{\mathcal{OBC}}]=\det[E-\mathcal{\widetilde{H}}_{\mathcal{OBC}}]$, one can prove that
{\it if $\det[E_{\mathcal{OBC}}-\mathcal{H}_{\mathcal{OBC};~n-2\times n-2}]=0$, then $\det[E_{\mathcal{OBC}}-\mathcal{H}_{\mathcal{OBC};~n\times n}]\approx0$}.
Although $\mathcal{\widetilde{H}}_{\mathcal{OBC},n\times n}$ has two additional eigenvalues compared with $\mathcal{\widetilde{H}}_{\mathcal{OBC},n-2\times n-2}$, such an approximation can still be regarded as valid.

\subsection{E. Deviation of Eq. (\ref{EQS43})}

For simplicity, if $t\pm \gamma + x>0$, $\forall$ $x\in[-\frac{W}{2},\frac{W}{2}]$, then one has [considering $\int \ln(a+bx)dx=(x+\frac{a}{b})[\ln(a+bx)-1)$]
\begin{align}
\begin{split}
\ln(\widetilde{\beta})=&\int_{-W/2}^{W/2}\ln[t-\gamma+x]\frac{dx}{2W}
-\int_{-W/2}^{W/2}\ln[t+\gamma+x]\frac{dx}{2W}\\
=&\frac{1}{2W}\{(t-\gamma+x)[\ln(t-\gamma+x)-1]|_{-W/2}^{W/2}
-(t+\gamma+x)[\ln(t+\gamma+x)-1]|_{-W/2}^{W/2}\}\\
=&\frac{1}{2W}\{(t-\gamma+\frac{W}{2})[\ln(t-\gamma+\frac{W}{2})-1]
-(t-\gamma-\frac{W}{2})[\ln(t-\gamma-\frac{W}{2})-1]\\
&~~~ -(t+\gamma+\frac{W}{2})[\ln(t+\gamma+\frac{W}{2})-1]
+(t+\gamma-\frac{W}{2})[\ln(t+\gamma-\frac{W}{2})-1]\}\\
=&\frac{1}{2W}\{(t-\gamma+\frac{W}{2})\ln(t-\gamma+\frac{W}{2})
-(t-\gamma-\frac{W}{2})\ln(t-\gamma-\frac{W}{2})\\
&~~~ -(t+\gamma+\frac{W}{2})\ln(t+\gamma+\frac{W}{2})
+(t+\gamma-\frac{W}{2})\ln(t+\gamma-\frac{W}{2})\}.\\
\end{split}
\end{align}

Importantly, we still have to prove that Eq. (\ref{EQS43}) is well-defined over the entire parameter region of $(W,\gamma)$, especially when
\begin{align}
\begin{split}
t\pm \gamma + x\le 0.
\end{split}
\end{align}
The key point is to solve the integral $\int_{b}^a\ln(x)dx$ when $a>0$ and $b<0$. Since $\ln[x]|_{x=0}$ is not well-defined and has no lower boundary, the {\it improper integral} instead the usual integral has to be adopted. Nevertheless, these two processes give the same result, which verifies the validity of Eq. (\ref{EQS43}) over the entire parameter region. The proof is given as follows as the integral $\int_{b}^a\ln(x)dx$ can be rewritten as:
\begin{align}
\begin{split}
\int_{b}^a\ln(x)dx&=\lim_{\varepsilon\rightarrow0}
\int_{b}^{-\varepsilon}\ln(x)dx+\lim_{\varepsilon\rightarrow0}\int^a_{\varepsilon}\ln(x)dx\\
&=\lim_{\varepsilon\rightarrow0}[x\ln(x)-x]|_{b}^{-\varepsilon}
+\lim_{\varepsilon\rightarrow0}[x\ln(x)-x]|^{a}_{\varepsilon}\\
&=[x\ln(x)-x]|_{b}^a+\lim_{\varepsilon\rightarrow0}
\{[-\varepsilon\ln(-\varepsilon)+\varepsilon]-[\varepsilon\ln(\varepsilon)-\varepsilon] \}\\
&=[x\ln(x)-x]|_{b}^a.
\end{split}
\end{align}
The relation $\lim\limits_{x\rightarrow0}[\pm x\ln(\pm x)]=0$ has been inserted in the final step. Thus, Eq. (\ref{EQS43}) is well-defined over the entire parameter space, except for the points where the denominator equals to zero. Fortunately, these points are strongly limited. Thus, it is appropriate to claim that Eq. (\ref{EQS43}) is well-defined over the entire parameter region.

\subsection{F. Analytical formula for the double-chain cases with $t_y\rightarrow\infty$: deviation of Eq. (\ref{EQbetad})}

In this part, we give the deviations of the analytical formula for the double-chain cases in Fig. \ref{Re2_2}(a).
We start from the following equation:
\begin{equation}
E-\mathcal{H}_{y;\mathcal{PBC};~N\times N} =
\left(
\begin{array}{cccccccccccccccc}
E   & t_{y;1}^+ & 0   & 0   & 0    & 0   & 0 & 0        & \cdots   & t_{y;L}  \\
t_{y;1}^- & E   & t_{y;2}^+ & 0   & 0    & 0   & 0 & 0        & \cdots   & 0     \\
0   & t_{y;2}^- & E   & t_{y;3}^+ & 0    & 0   & 0 & 0        & \cdots   & 0   \\
0   & 0   & t_{y;3}^- & E   & t_{y;4}^+  & 0   & 0 & 0        & \cdots   & 0   \\
0   & 0   & 0   & t_{y;4}^- & E    & t_{y;5}^+ & 0 & 0        & \cdots   & 0  \\
0   & 0   & 0   & 0   & t_{y;5}^- & E   & t_{y;6}^+  & 0        & \cdots   & 0     \\
\vdots   & \vdots   & \vdots   & \vdots   & \vdots    & \ddots   & \ddots & \ddots        & \ddots   & 0   \\
0   & 0   & 0   & 0   & 0    & 0   & 0 & E        & t_{y;n-2}^+ & 0   \\
0   & 0   & 0   & 0   & 0    & 0   & 0 & t_{y;n-2}^- & E        & t_{y;N-1}^+   \\
t_{y;R}   & 0   & 0   & 0   & 0    & 0   & 0 & 0        & t_{y;N-1}^-   & E   \\
\end{array}
\right)_{N\times N}.
\label{EQhn}
\end{equation}
we set $y\in[a,b]$. $a$ and $b$ stand for two Hatano-Nelson chains with hopping strength marked in the equation. $t_{a/b;L}$ and $t_{a/b;R}$ ensures the periodic boundary conditions(PBC). $E$ is the eigenvalue. The Hamiltonian for the double-chain case satisfies:
\begin{equation}
E-\mathcal{H}_{d; \mathcal{PBC};~2N\times 2N}=
\left(
\begin{array}{cc}
E-\mathcal{H}_{a;\mathcal{PBC};~N\times N} & t_y\mathbb{I}_{N\times N} \\
t_y\mathbb{I}_{N\times N} & E-\mathcal{H}_{b;\mathcal{PBC};~N\times N}
\end{array}
\right)_{2N\times 2N},
\end{equation}
with $\mathbb{I}_{N\times N}$ the unitary matrix. The determinant satisfies:
\begin{equation}
\det[E-\mathcal{H}_{d; \mathcal{PBC};~2N\times 2N}]=\det[(E-\mathcal{H}_{a;\mathcal{PBC};~N\times N})(E-\mathcal{H}_{b;\mathcal{PBC};~N\times N})-t_y^2\mathbb{I}_{N\times N}].
\end{equation}
Then, $(E-\mathcal{H}_{a;\mathcal{PBC};~N\times N})(E-\mathcal{H}_{b;\mathcal{PBC};~N\times N})$ gives raise to
\begin{equation}
\begin{array}{l}
(E-\mathcal{H}_{a;\mathcal{PBC};~N\times N})(E-\mathcal{H}_{b;\mathcal{PBC};~N\times N})-t_y^2\mathbb{I}_{N\times N}=\\
\left(
\begin{array}{cccccccccccccccc}
C_1   & E(t_{a;1}^++t_{b;1}^+) & t_{a,1}^+t_{b;2}^+   & 0   & 0    & 0   & 0 & \cdots        & t_{a;L}t^-_{b;N-1}   & E(t_{a;L}+t_{b;L})  \\
E(t_{a;1}^-+t_{b;1}^-) & C_2   & E(t_{a;2}^++t_{b;2}^+) & t_{a,2}^+t_{b;3}^+   & 0    & 0   & 0 & \cdots       & 0   & t_{b;L}t^-_{a;1}     \\
t_{a,1}^-t_{b;2}^-   & E(t_{a;2}^-+t_{b;2}^-) & C_3   & E(t_{a;3}^++t_{b;3}^+) & t_{a,3}^+t_{b;4}^+    & 0   & 0 & \cdots        & \cdots   & 0   \\
\vdots   & \vdots   & \vdots   & \vdots   & \vdots    & \ddots   & \ddots & \ddots        & \cdots   & \vdots   \\
0   & 0   & 0   & 0   & 0    & 0   & \ddots & \ddots        & \cdots & 0   \\
t_{b;R}t^+_{a;N-1}   & 0   & 0   & 0   & 0    & 0   & 0 &\ddots & \cdots        & E(t_{a;N-1}^++t_{b;N-1}^+)   \\
E(t_{a;R}+t_{b;R})   & t_{a;R}t^+_{b;1}   & 0   & 0   & 0    & 0   & 0 & \cdots        & \ddots   & C_{N}   \\
\end{array}
\right)_{N\times N}.
\end{array}
\end{equation}
We mark $C_n=E^2-t_y^2+(t_{a;n}^+t_{b;n}^-+t_{a;n-1}^+t_{b;n-1}^-)$. Specifically, one has $C_1=E^2-t_y^2+(t_{a;1}^+t_{b;1}^-+t_{a;L}t_{b;R})$ and $C_{N}=E^2-t_y^2+(t_{a;N-1}^+t_{b;N-1}^-+t_{a;R}t_{b;L})$.

Based on our theory, one needs to calculate all the terms correlated with $t_{a/b;L/R}$. However, the complicated matrix formula for $[(E-\mathcal{H}_{a;\mathcal{PBC};~N\times N})(E-\mathcal{H}_{b;\mathcal{PBC};~N\times N})-t_y^2\mathbb{I}_{N\times N}]$ makes it almost impossible. Nevertheless, the problem can be simplified by considering the following two points.\\
(1). For $N\rightarrow \infty$, all the eigenvalues roughly sitting at $E=\pm t_y+\delta$. $\delta$ is correlated to the value of $t^{\pm}_{a/b;i}$. \\
(2). For $|t_y|\gg |t^\pm_{a/b;i}|$, $\delta$ can be neglected compared with $t_y$.\\
Thus, one only needs to calculate the transformation parameter $\beta$ for $E\approx\pm t_y$. By requiring $E=t_y$, $\det[(E-\mathcal{H}_{a;\mathcal{PBC};~N\times N})(E-\mathcal{H}_{b;\mathcal{PBC};~N\times N})-t_y^2\mathbb{I}_{N\times N}]$ can be rewritten as:
\begin{equation}
\begin{array}{l}
\det[(t_y-\mathcal{H}_{a;\mathcal{PBC};~N\times N})(t_y-\mathcal{H}_{b;\mathcal{PBC};~N\times N})-t_y^2\mathbb{I}_{N\times N}]=\\
\left|
\begin{array}{cccccccccccccccc}
t_{a;1}^+t_{b;1}^-+t_{a;L}t_{b;R}   & t_y(t_{a;1}^++t_{b;1}^+) & t_{a,1}^+t_{b;2}^+   & 0   & 0    & 0   & 0 & \cdots        & t_{a;L}t^-_{b;N-1}   & t_y(t_{a;L}+t_{b;L})  \\
t_y(t_{a;1}^-+t_{b;1}^-) & \sum_{i=1,2}(t_{a,i}^+t_{b;i}^-)   & t_y(t_{a;2}^++t_{b;2}^+) & t_{a,2}^+t_{b;3}^+   & 0    & 0   & 0 & \cdots       & 0   & t_{b;L}t^-_{a;1}     \\
t_{a,1}^-t_{b;2}^-   & t_y(t_{a;2}^-+t_{b;2}^-) & \ddots  & t_y(t_{a;3}^++t_{b;3}^+) & t_{a,3}^+t_{b;4}^+    & 0   & 0 & \cdots        & \cdots   & 0   \\
\vdots   & \vdots   & \vdots   & \vdots   & \vdots    & \ddots   & \ddots & \ddots        & \cdots   & \vdots   \\
0   & 0   & 0   & 0   & 0    & 0   & \ddots & \ddots        & \cdots & 0   \\
t_{b;R}t^+_{a;N-1}   & 0   & 0   & 0   & 0    & 0   & 0 &\ddots & \cdots        & t_y(t_{a;N-1}^++t_{b;N-1}^+)   \\
t_y(t_{a;R}+t_{b;R})   & t_{a;R}t^+_{b;1}   & 0   & 0   & 0    & 0   & 0 & \cdots        & \ddots   & \ddots   \\
\end{array}
\right|.
\end{array}
\end{equation}
Considering $|t_y|\gg |t^\pm_{a/b;i}|$, one only needs to consider terms correlated with $t_y$. The above equation can be simplified as follows:
\begin{equation}
\begin{array}{l}
\det[(t_y-\mathcal{H}_{a;\mathcal{PBC};~N\times N})(t_y-\mathcal{H}_{b;\mathcal{PBC};~N\times N})-t_y^2\mathbb{I}_{N\times N}]\approx\\
\left|
\begin{array}{cccccccccccccccc}
0   & t_y(t_{a;1}^++t_{b;1}^+) & 0   & 0   & 0    & 0   & 0 & \cdots        & 0   & t_y(t_{a;L}+t_{b;L})  \\
t_y(t_{a;1}^-+t_{b;1}^-) & 0   & t_y(t_{a;2}^++t_{b;2}^+) & 0   & 0    & 0   & 0 & \cdots       & 0   & 0     \\
0   & t_y(t_{a;2}^-+t_{b;2}^-) & 0   & t_y(t_{a;3}^++t_{b;3}^+) & 0    & 0   & 0 & \cdots        & \cdots   & 0   \\
\vdots   & \vdots   & \vdots   & \vdots   & \vdots    & \ddots   & \ddots & \ddots        & \cdots   & \vdots   \\
0   & 0   & 0   & 0   & 0    & 0   & \ddots & \ddots        & \cdots & 0   \\
0   & 0   & 0   & 0   & 0    & 0   & 0 &\ddots & \cdots        & t_y(t_{a;N-1}^++t_{b;N-1}^+)   \\
t_y(t_{a;R}+t_{b;R})   & 0   & 0   & 0   & 0    & 0   & 0 & \cdots        & \ddots   & 0   \\
\end{array}
\right|_{N\times N}.
\end{array}
\end{equation}
It is similar to the one-chain cases, except the hopping is changed to $t_y(t_{a;i}^\pm+t_{b;i}^\pm)$. Based on our previous deviations, the minimum correlated to $t_{a/b;L/R}$ is determined by the following equations:
\begin{equation}
\mathcal{F}(\beta)\approx | \chi\prod_{i=1}^{N-1} [t_y^N\beta^{-N}(t_{a;i}^- + t_{b;i}^-)(t_{a;L}+t_{b;L})]+
\chi\prod_{i=1}^{N-1}[t_y^N\beta^{N}(t_{a;i}^+ + t_{b;i}^+)(t_{a;R}+t_{b;R})]|.
\end{equation}
$\chi$ is a constant.
 These terms are the largest leading term balancing the influence of $\beta$ and $t_y$.
Then, the required minimum gives:
\begin{equation}
\widetilde{\beta}_d^{2N}\approx \prod_{i=1}^{N} [\frac{t_y^N(t_{a;i}^- + t_{b;i}^-)}{t_y^N(t_{a;i}^+ + t_{b;i}^+)}]=\prod_{i=1}^{N} [\frac{2t^- + w_{a;i}^- + w_{b;i}^-}{2t^+ + w_{a;i}^+ + w_{b;i}^+}].
\end{equation}
Although we require $|t_y|\gg |t^\pm_{a/b;i}|$ in our deviation, the final result is still an highly accurate approximation when $|t_y|> |t_{a,b;i}^{\pm}|$. Such a result looks similar to the Hatano-Nelson model. However, they show distinct behaviors since $w_{a;i}^\pm$ and $w_{b;i}^\pm$ are independent to each other.

\subsection{ G. Self-consistent Born approximation method could be unreliable for disordered non-Hermitian systems }

Following previous study in non-Hermitian systems \cite{disS11}, the self-consistent Born
approximation (SCBA) is implemented.
 Since self-energy of SCBA is independent of momentum \cite{SCBAS1}, we pay our attention to the following Hamiltonian
\begin{equation}
H=H_0(k_x)+V_{dis}=
\left(
\begin{array}{cc}
0 & t+t_0e^{ik_x}\\
t+t_0e^{-ik_x} &0
\end{array}
\right)+
\left(
\begin{array}{cc}
0 & dis\\
0 &0
\end{array}
\right).
\label{SCBAMODEL}
\end{equation}
$dis$ stands for the disorder.
We set $t_0=1$ and $t=1$. Such a Hamiltonian has the same features ($\beta$) as the Hamiltonian adopted in the main text [based on Eq. (8) in the main text].
One has:
\begin{equation}
 G=g_0+g_0V_{dis}g_0+g_0V_{dis}g_0V_{dis}g_0+\cdots,
\end{equation}
where $G=[E-H]^{-1}$ and $g_0=[E-H_0(k_x)]^{-1}$.
The self energy is \cite{disS11}:
\begin{align}
\begin{split}
\Sigma=&\Sigma_1+\Sigma_2+\Sigma_3+\cdots\\
=&\langle V_{dis}\rangle+\langle V_{dis}g_0V_{dis}\rangle+ \langle V_{dis}g_0V_{dis}g_0V_{dis}\rangle+\cdots
\end{split}
\end{align}
where $\Sigma_2=\langle V_{dis}g_0V_{dis}\rangle$. To contain the influence of $\Sigma_n$ with $n\in\mathbb{Z}$, the SCBA is implemented as follow \cite{SCBAS1}:
\begin{equation}
\Sigma_2=\frac{\langle dis^2\rangle}{2\pi}M\int_{-\pi}^{\pi}[E+i\eta-H_0(k_x)-\Sigma_1-\Sigma_2]^{-1}dk_xM=\left(
\begin{array}{cc}
0&\delta_2\\
0&0
\end{array}
\right),
\label{SCBA1}
\end{equation}
with $M=\left(
\begin{array}{cc}
0&1\\
0&0
\end{array}
\right).
$
We mark $\Sigma_1=\left(
\begin{array}{cc}
0&\delta_1\\
0&0
\end{array}
\right)$ with $\delta_1=\langle dis\rangle$.

By setting $w\in[-W/2,W/2]$, two typical cases are considered:\\
(1) For the disorder case $(t+w)$ , one has $dis=w$. Thus, one has $\delta_1=\langle dis\rangle=0$ and $\langle dis^2\rangle=\frac{W^2}{12}$.\\
(2) For the disorder case $te^w$, one has $dis=(te^w-t)$. By considering $t=1$, one has
\begin{align}
\begin{split}
\delta_1&=\langle dis\rangle =\int_{-W/2}^{W/2}\frac{1}{W}(e^w-1)dw=(e^{W/2}-e^{-W/2}-W)/W
\end{split}
\end{align}
 and
 \begin{align}
\begin{split}
\langle dis^2\rangle&=\int_{-W/2}^{W/2}\frac{1}{W}(e^w-1)^2dw=[\frac{e^W-e^{-W}}{2}-2(e^{W/2}-w^{-W/2})+W]/W.
\label{SCBA2}
\end{split}
\end{align}
The self-energy only gives the renormalization of the Hamiltonian's element $H(1,2)$, among which $t\rightarrow T=t+\delta_1+\delta_2$.

 By considering $E=0$ and $\eta\rightarrow0^+$,  Eq. (\ref{SCBA1}) can be rewritten as:
\begin{align}
\begin{split}
\Sigma_2=\left(
\begin{array}{cc}
0&\delta_2\\
0&0
\end{array}
\right)&=\frac{\langle dis^2\rangle}{2\pi}\int_{-\pi}^{\pi}M[E+i\eta-H_0(k_x)-\Sigma_1-\Sigma_2]^{-1}Mdk_x\\
&=-\frac{\langle dis^2\rangle}{2\pi}\int_{-\pi}^{\pi}M\left(
\begin{array}{cc}
-i\eta&t+t_0e^{ik_x}+\delta_1+\delta_2\\
t+t_0e^{-ik_x}&-i\eta
\end{array}
\right)^{-1}Mdk_x
\\
&\approx
\left(
\begin{array}{cc}
0&-\frac{\langle dis^2\rangle}{2\pi}\int_{-\pi}^{\pi}(T+t_0e^{ik_x})^{-1}dk_x\\
0&0
\end{array}
\right)臓贈
\end{split}
\end{align}
Unfortunatly, the self-consistent process may not be available to achieve a reliable results for the considered models. Taking the first case as an example, we notice the correct results require $T^2<t_0^2=1$ for $W<5$ [see Fig.3 in the main text]. Put the correct results into the equation, one will have $\delta_2=-\frac{\langle dis^2\rangle}{2\pi}\int^\pi_{-\pi}(T+e^{ik_x})^{-1}dk_x\approx0$ since $\int_{-\pi}^\pi\frac{dk_x}{T+e^{ik_x}}=0$ when $|T|<1$.
 Thus, each time one tends to approach the correct results, the self-consistent process will reset $\delta_2\rightarrow0$. In other words, $\delta_2\rightarrow 0^-$ is roughly the solution of the self-consistent equation for the first case. However, such a result is inconsistent with the correct results in the main text. Similar discussions are also available by considering $\delta_1\neq0$ for the second case.

\begin{figure}[t]
   \centering
    \includegraphics[width=0.6\textwidth]{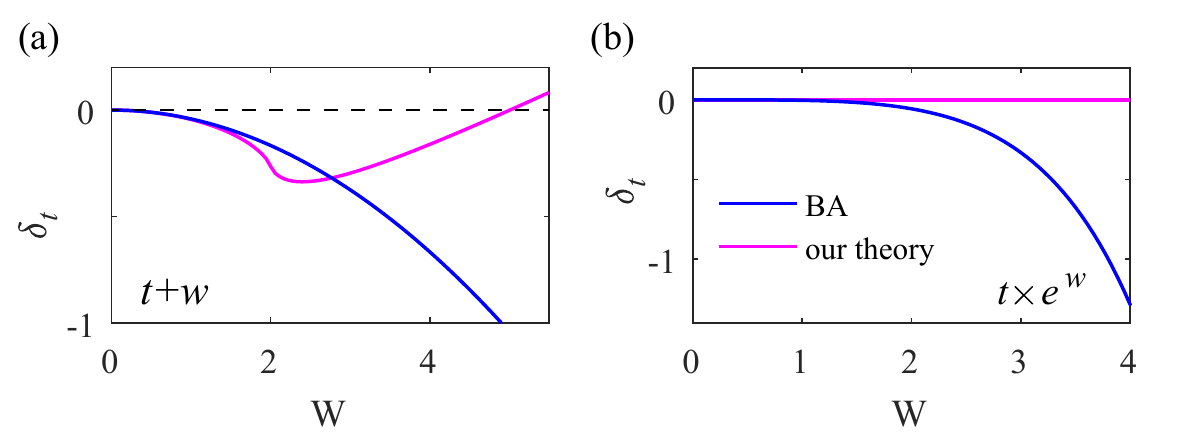}
    \caption{(Color online). The renormalization of the hopping strength $\delta_t=\delta_1+\delta_2$ based on Born-approximation for disorder schemes: (a) $t+w$ and (b) $t\times e^w$. The disorder satisfies $w\in[\frac{-W}{2},\frac{W}{2}]$. The pink solid lines are obtained by our theory with $\delta_t=|\widetilde{\beta}|^2t-t$, where $\widetilde{\beta}^2=(t+\delta_t)/t$. $\widetilde{\beta}$ has been given in the main text. We set $t=1$, $t_0=1$ and $E=0$ in our calculations.  The NHSE is absent if $\delta_t=\delta_1+\delta_2=0$. For $\delta_t\neq0$, the direction of NHSE is determined by the sign of $\delta_t$.     }
   \label{Re6}
\end{figure}

Based on these considerations, we pay our attention to the Born-approximation (BA) $\delta_t=\delta_1+\delta_2$ analytically. Such a process is also widely used in the study of disorder-induced renormalization in Hermitian systems \cite{SCBAS1}.
 One has
\begin{equation}
\delta_2=-\frac{\langle dis^2\rangle}{2\pi}\int^\pi_{-\pi}(1+e^{ik_x})^{-1}dk_x=-\frac{\langle dis^2\rangle}{2}.
\end{equation}
 As shown in Fig. \ref{Re6}(a), the BA captures the existence of disorder-induced NHSE for disorder-enhanced NHSE conditions. Nevertheless, the value of $\delta_t$ is inconsistent with our theory when $W>1.5$,
and the disorder-induced reversing of NHSE can not be observed through the proposed BA approach [see Fig. \ref{Re6}(a)]. As for the disorder-irrelevant NHSE cases, the BA fails, where the BA predicts the existence of NHSEs for such a case, as shown in Fig. \ref{Re6}(b).

On the theoretical side,
 the BA and SCBA are taking some kinds of algebraic average of the disorder [see Eqs. (\ref{SCBA1})-(\ref{SCBA2})]. However, the correct results require the geometric average of the disorders, and this is what we do in our work.
  Based on the above studies, it is reasonable to conclude that the SCBA is not a valid approach to describe the NHSEs in disordered non-Hermitian systems.

\end{widetext}


\begin{thebibliography}{99}
\bibitem{footnote3} Since $\mathcal{H}$ and $\mathcal{\widetilde{H}}$ have the same open boundary spectrum, they should have the same topological features and phase diagrams. This explains why GBZ theory can capture the bulk-boundary correspondences in non-Hermitian systems.
\bibitem{footnote5} For GBZ theory, the first issue is naturally ensured. The second issue is available by ensuring the extended wavefunction distributions of $\mathcal{\widetilde{H}_{OBC}}$ \cite{NHSE1}. It suggests $\mathcal{\widetilde{H}}$ has no NHSE, consisting with our theory.
\bibitem{footnote4} The self-consistent Born approximation is taking some kinds of algebraic average for disorder \cite{SCBA1}. The correct results for NHSEs require the geometric average for the disorders. Thus, the SCBA is not a valid approach to describe the NHSEs in disordered non-Hermitian systems [see SM. VII for more details].
\bibitem{footnote7} We emphasize that $\widetilde{\beta}$ is not always the geometric mean of $\beta_i=\sqrt{t_i^-/t_i^+}$. It is true only for the Hatano-Nelson model.
\bibitem{footnote6} The $\mathcal{PBC}$ still holds for the modified GBZ theory since the GBZ theory with $\beta_i\equiv\widetilde{\beta}\rightarrow\beta_{\mathcal{GBZ}}$ gives the wrong eigenvalues.
\bibitem{footnote7} We found there exists a critical $t_y$ between $\widetilde{\beta}_{a/b}>1$ and $\widetilde{\beta}_{d}>1$ with $t_y=t_c\approx2.8$ for the considered model and parameters. The coupling-reversed NHSE's is available when $t_y>2.8$. Such a value is consistent with the requirement $|t_y|>|t^\pm_{y;i}|$.
\end{thebibliography}
\end{document}